\documentclass{article}

\usepackage[english]{babel}

\usepackage[letterpaper,top=2cm,bottom=2cm,left=3cm,right=3cm,marginparwidth=1.75cm]{geometry}

\usepackage{amsmath}
\usepackage{lipsum}
\usepackage[dvipsnames]{xcolor}
\usepackage{graphicx}
\usepackage{cite} % to group consecutive citations
\usepackage[colorlinks=true, allcolors=blue]{hyperref}

\newcommand{\x}{{\mathbf{x}}}

\newcommand{\Jij}{{J}_{ij}}
\newcommand{\rr}{{\mathbf{r}}}
\newcommand{\kk}{\mathbf{k}}
\newcommand{\F}{\mathcal{F}}
\newcommand{\eg}{\textit{e.g.} }
\newcommand{\qea}{q_\mathrm{EA}}

\graphicspath{{./figures/}{./}}

\title{An Introduction to the Theory of Spin Glasses}
\author{Ada Altieri$^1$ \& Marco Baity-Jesi$^2$}
\date{%
    $^1$ Laboratoire Mati\`ere et Syst\`emes Complexes (MSC), Universit\'e Paris Cit\'e
CNRS, 75013 Paris, France\\%
    $^2$Eawag (ETH), \"Uberlandstrasse 133, CH-8600 D\"ubendorf, Switzerland\\[2ex]%
}

\begin{document}
\maketitle

\begin{abstract}
We review the main methods used to study spin glasses. In the first part, we focus on methods for fully connected models and systems defined on a tree, such as the replica method, the Thouless-Anderson-Palmer formalism, the cavity method, and the dynamical mean-field theory. In the second part, we deal with the description of low-dimensional systems, mostly in three spatial dimensions, which are mostly studied through numerical simulations. 
We conclude by mentioning some of the main open problems in the field.
\end{abstract}

\tableofcontents

\newpage

\section{Spin Glasses}
Spin glasses are paradigmatic complex systems, for which disorder plays a central role. 
Although the term was originally coined to describe certain magnetic materials that exhibit an exotic phase behavior, associated models and theory have since found applications in a wide variety of fields, thus making the study of spin glasses an intrinsically interdisciplinary endeavor.

Here, we provide an overview of spin glasses (SGs), from their experimental context to their common theoretical models. We notably present several theoretical methods developed within the context of their study, describe the differences between mean-field and three-dimensional SGs, discuss various interdisciplinary applications, and introduce some open questions in the field.\\

% Spin glasses (SGs) are a paradigmatic complex system, where disorder plays a central role. 
% If on one side, their relevance is due to their exotic behavior, on another side, spin glass models and theory are used throughout a wide variety of fields, automatically making the study of spin glasses an intrinsically interdisciplinary topic.

% Here, we provide a description of spin glasses, from the experimental system, to the spin-glass models that are used in interdisciplinary applications. We introduce several methods that were developed within the context of spin glass theory, discuss the differences between mean-field and three-dimensional SGs, and point at interdisciplinary applications and open questions. \\

\paragraph{Experimental SGs}
As materials, SGs are disordered magnetic alloys containing strongly interacting ions immersed in a weakly interacting {substrate~\cite{mydosh:93,binder:86, nordblad:16, Vincent:22}. They are prepared by rapidly cooling the liquid alloy, thus fixing the strongly interacting particles at random positions within the resulting solid. The pairwise exchange interaction between ions is then positive or negative, depending on the distance vector $\rr$ between ions, such as they are for {Ruderman-Kittel-Kazuya-Yosida (RKKY)} interactions~\cite{ruderman:54,kasuya:56,yosida:57},
\begin{equation}\label{eq:rkky}
    J(\rr)\sim \frac1{|\rr|^3}\cos(\kk\cdot\rr)\,,
\end{equation}
where the modulus of the frequency $\kk$ is of the order of the Fermi vector. Spin glasses also arise in systems with interactions different from RKKY. The general idea is that because ion positions depend on the specific realization of the alloy, distances between them are randomly distributed (and a priori unknown), and therefore values of $J(\rr)$ are randomly positive and negative. 

To make these systems more physically tractable, we can define SG models assuming that the distances between ions are fixed (for example, on a lattice), and that the coupling between two ions -- commonly called spins -- is a quenched random variable~\cite{anderson:70}. \textit{Quenched} variables here refer to random quantities that appear in the Hamiltonian as parameters that do not change during the evolution of the system, so as to capture that ions' positions are fixed over the relevant experimental time scales.
This formulation leads to the generic SG Hamiltonian
\begin{equation}\label{eq:sg}
    \mathcal{H} = -\sum_{i<j}^N s_i \cdot \Jij s_j\,,
\end{equation}
where the $N$ dynamical variables are spins $s_i$ coupled through pairwise interactions of quenched magnitude $\Jij$. 
Spins can be formulated in different ways, but in this section we restrict our consideration to Ising spins, $s_i=\pm1$, both because of their simplicity, and because experimental systems are typically spatially anisotropic~\cite{dzyaloshinsky:58,moriya:60,fert:80}, thus rendering effective interactions Ising-like~\cite{martin-mayor:11b,baityjesi:14}.
The quenched random variables are extracted from a distribution $P(\Jij)$, which has support on both positive and negative values. In fully-connected models, all couplings are extracted from $P(\Jij)$, while in other models some of the couplings are set to zero. For example, the Edwards-Anderson model (EAM) is defined by Hamiltonian~\eqref{eq:sg} on a $d$-dimensional square lattice, so if $s_i$ and $s_j$ are not nearest neighbors then $\Jij$ is suppressed.

Although Eq.~\eqref{eq:sg} is not the only way to model SGs (for example, interactions can be more than pairwise), two of its features are nevertheless generic: quenched disorder and frustration~\cite{young:98}. 
Frustration here refers to the impossibility to satisfy simultaneously all local constraints, which follows from the couplings being i.i.d. variables that can be both positive and negative. As illustrated in Fig.~\ref{fig:frustration}, along a loop local couplings cannot all be minimized.

\begin{figure}[t]
    \centering
    \includegraphics[width=0.37\textwidth]{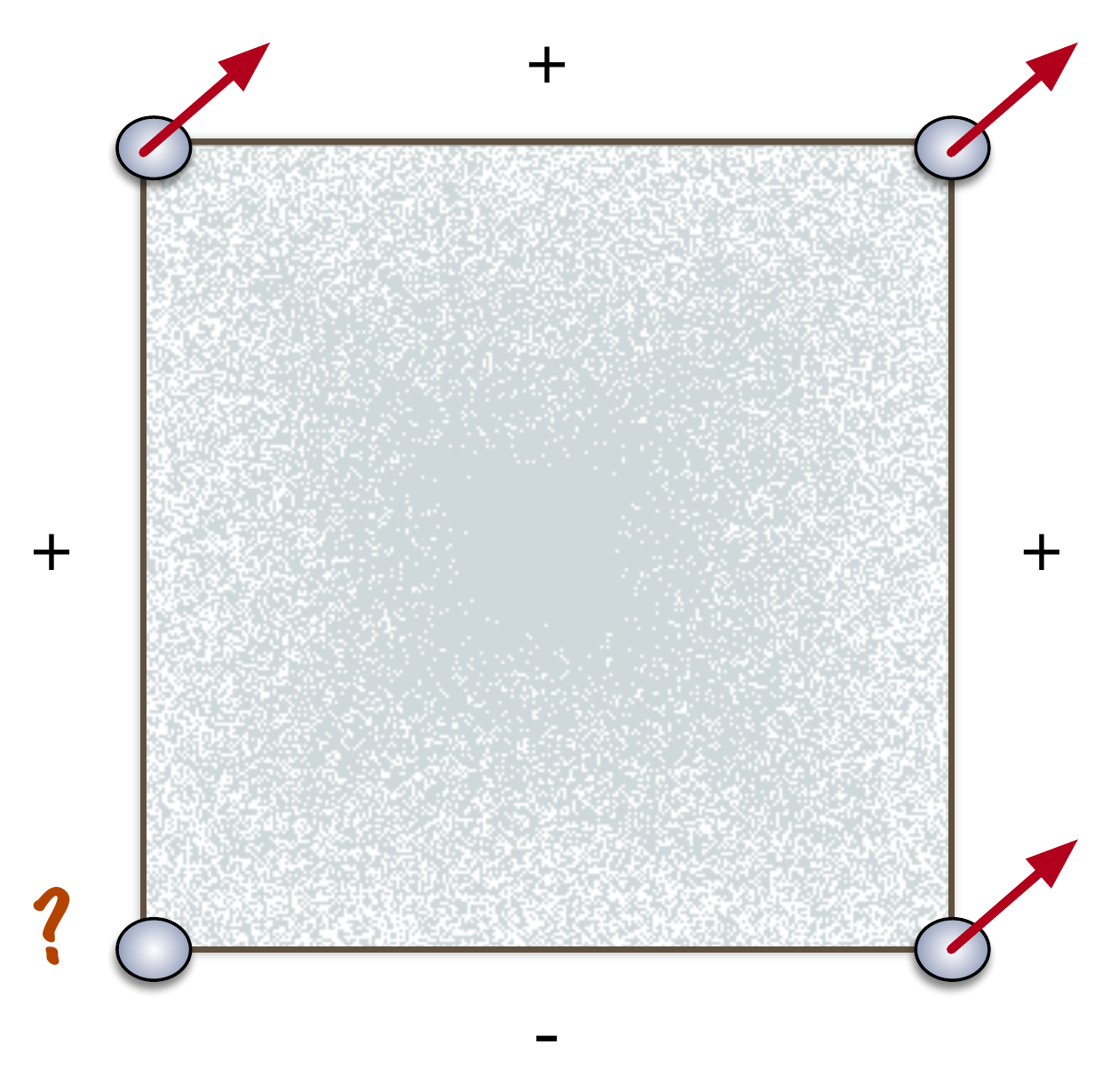}
    \caption{Disordered interactions result in frustration. The signs along the edges indicate whether two neighboring spins prefer to point in the same (+) or opposite (-) directions. It is therefore impossible to satisfy all the couplings simultaneously on the square plaquette.}
    \label{fig:frustration}
\end{figure}

\paragraph{Self-averaging} Because of the quenched disorder, every SG Hamiltonian is different from all others. In other words, every \textit{sample} corresponds to a different set of couplings $\Jij$. Interestingly, even though samples are microscopically different, experimental SG realizations display the same macroscopic behavior. SG descriptions thus work under the assumption, which holds in most relevant cases~\cite{guerra:02,carmona:06,chatterjee:05,jagannath:22}, that large SG samples display equivalent average behavior.
Therefore, although a \textit{quenched partition function}, $Z_J$, can be computed for each sample, the quantity of physical interest is the \textit{quenched free energy} averaged $\overline{(\ldots)}$ over the couplings distribution,
\begin{equation}\label{eq:F}
    \F = \overline{F_J} = -T\overline{\log Z_J},
\end{equation}
for Boltzmann constant set to unity $k_\mathrm{B}=1$, temperature $T$.

\paragraph{Order parameter}
The SG order parameter is also an important matter. In a ferromagnetic system, magnetization, $M=\tfrac{1}{N}\langle\sum_i s_i\rangle$, distinguishes between the ferromagnetic and the paramagnetic phases. However, for the Hamiltonian in Eq.~\eqref{eq:sg} -- if $P(\Jij)$ is symmetric around zero -- the magnetization is zero for all temperatures.

% {\color{red} Unlike ferromagnetic systems,
% states cannot be classified just in terms of their symmetry. In principle, to select a state, one should introduce an infinitesimal local magnetic field different for each spin variable. However, this is not enough to solve the problem: since the local magnetization depends on the disordered couplings, the infinitesimally small field must be applied before averaging over the disorder, which in turn leads to unavoidable correlations between the local field and the quenched disorder.
% }
If a low-temperature SG phase exists, %however, although this will not be visible through $M$, 
there must nevertheless be some configurations that are preferred over 
% this means that one or more spin configurations will be preferred to 
others. One way of identifying these preferred configurations is through the \textit{Edwards-Anderson overlap}~\cite{edwards:75,edwards:76},
\begin{equation}\label{eq:qea}
    \qea = \frac1N\lim_{t\to\infty} \sum_{i=1}^{N} \left\langle s_i(0) s_i(t) \right\rangle_t\,,
\end{equation}
where $\langle(\ldots)\rangle_t\equiv\frac{1}{t}\int_0^t (\ldots)dt$ marks a time average.
If spins $s_i$ have no preferred value, as in the paramagnetic phase, then the products in Eq.~\eqref{eq:qea} vanish on average, and $\qea=0$;  but if some configurations are preferred, each spin has a preferential value, and $\qea>0$.

As further discussed in Sec.~\ref{sec:replica}, the overlap can also be expressed without resorting to time averages, through the concept of replicas. Replicas of a system have exactly the same couplings $J_{ij}$, but evolve independently. For $s_i^{(a)}$ and $s_i^{(b)}$ denoting spins that belong to replicas $a$ and $b$ of the same sample, we can then write the overlap as
\begin{equation}\label{eq:qab}
    q_{ab} = \frac1N\sum_{i=1}^{N} \left\langle s_i^{(a)}\cdot s_i^{(b)}\right\rangle\,,
\end{equation}
As above, the overlap will be zero if there is no preferred configuration, and positive otherwise.

\section{Analytical Methods for Spin Glasses}

Some of the most successful methods for studying SG Hamiltonians were developed to describe fully-connected models %(every spin interacts with any other spin) 
as well as models defined on tree-like graphs. The first geometry corresponds to the mean-field (MF) approximation, which becomes exact in the limit of $d\to\infty$ spatial dimensions, while the second goes beyond the MF description but neglects contributions from feedback loops.

\subsection{The Replica Method}\label{sec:replica}
At variance with the partition function, which is typically non-self-averaging, the free energy is key to probing the low-temperature behavior of SGs (as shown in Eq.~\eqref{eq:F}). However, calculating the average of the logarithm in Eq.~\eqref{eq:F} can be challenging.
The \emph{replica method} overcomes this difficulty by replacing the calculation of $\overline{\ln(Z)}$ with that of $\overline{Z^n}$ through the identity
\begin{equation}\label{eq:reptrick}
\overline{\ln{Z}}=\lim \limits_{n\rightarrow 0}
\frac{\overline{Z^{n}}-1}{n}\,,
\end{equation}
and by treating $Z^n$ as the partition function for $n$ replicas of the same sample. Treating the index $n$ as an integer, however, hinges on the assumption that the analytical continuation $n \rightarrow 0$ exists. Assuming it does, the replicated partition function can then be rewritten as a function of the overlap matrix $Q_{ab}$ describing the overlap between two replicas $a$ and $b$:
\begin{equation}
\overline{Z^{n}}=\int \prod_{(ab)} \; \frac{d Q_{ab}}{2\pi}
e^{N \mathcal{A}[Q_{ab}]} \ .
\end{equation}
Taking advantage of the thermodynamic limit $N \rightarrow \infty$, the above expression can be evaluated by the Laplace method of saddle-point approximation, which extremizes the action $\mathcal{A}$ with the respect to the reference order parameter. The final result, therefore, depends on the structure of $Q_{ab}$.

The simplest and most intuitive ansatz for that structure is the replica symmetric (RS) one~\cite{dealmeida:78,bray:78}. In this case, the overlap among different replicas is the same for any pair of replicas, with the exception of the overlap between a replica and itself. In the RS scenario, the matrix $Q_{ab}$ therefore admits only two values: a diagonal contribution $q_d$ (for $a=b$) and an off-diagonal one $q_0$ (for $a \neq b$).
As expected, the RS ansatz correctly describes the high temperature %{$T$}
and large external magnetic field 
regimes. But as these control parameters are lowered, some models reach a critical de Almeida-Thouless (dAT) line~\cite{dealmeida:78}, below which the RS solution becomes unstable. Physically, a \textit{spin glass} phase emerges when the RS solution is no longer stable. 

More technically, below the dAT line, the correct solution requires a matrix $Q_{ab}$ with an iterative block structure, which breaks the symmetry between different pairs of replicas~\cite{parisi:79,parisi:79b,parisi:80,parisi:80b,parisi:80c}. 
For the first iteration of replica symmetry breaking (noted 1-RSB), the $n \times n$ matrix is parametrized by a diagonal value $q_d$ and two off-diagonal values that can be either $q_1$, if the two replicas belong to the same block of size $m \times m$, or $q_0$, if the replicas fall outside the innermost block (see Fig.~\ref{1rsb_matrix}).
\begin{figure}[t]
    \centering
    \includegraphics[scale=0.52]{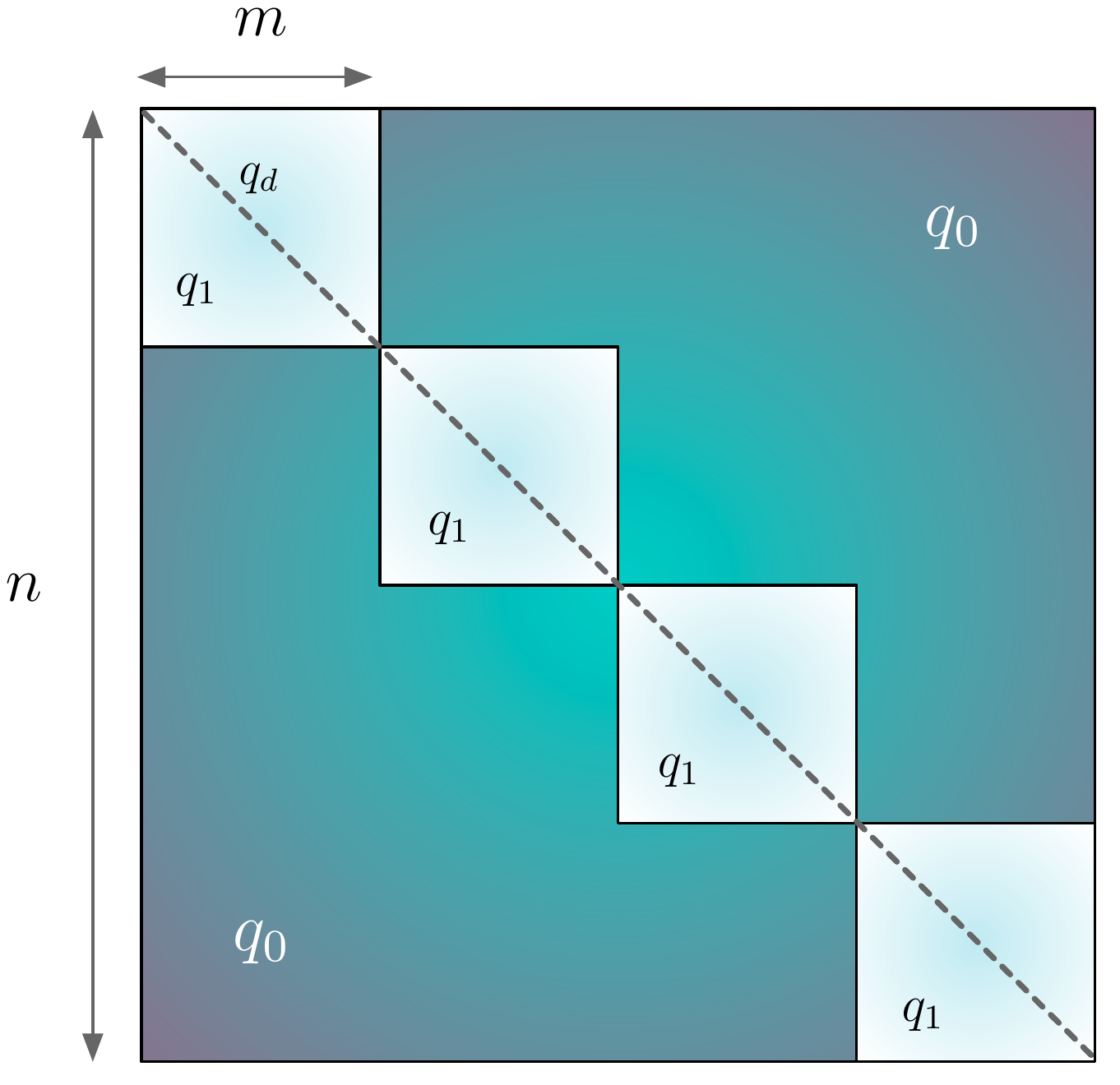}
    \caption{$1$-RSB parametrization of the Parisi overlap matrix. $q_1$ represents the degree of similarity between two replicas inside the innermost block of size $m \times m$, whereas $q_0$ is the outermost block value.}.
    \label{1rsb_matrix}
\end{figure}
If the low-temperature phase remains unstable after 1-RSB, this procedure can be iterated within each of these blocks, leading to a $k$-step RSB (or $k$-RSB).
Depending on the specifics of the Hamiltonian, SG phases can have different levels of RSB.
The limit $k \rightarrow \infty$ corresponds to the full-RSB scenario, according to which the overlap matrix is parametrized by a continuous function $q(x)$, with $x \in [0,1]$~\cite{parisi:80}. 
The overlap density distribution, $P(q)$, being non-zero over a continuous range of $q$, is a signature of the presence of an infinite number of symmetry-breaking points~\cite{mezard:87}.

% The continuous function $x(q)$ of breaking symmetry points is linked to the integral of the probability distribution of the overlap, $x(q)=\int_{0}^{q} dq'\; P(q')$, where $P(q)$ can also be extracted from large-deviation principles \cite{franz:95b}. 

In the replica formalism, all the mutual information about pairs of equilibrium configurations is encoded in the overlap, which is a measure of their mutual distance~\cite{ghirlanda:98, parisi:00origin}. Given the hierarchical way in which the full-RSB construction is obtained, in systems with a full-RSB phase, the states have an \textit{ultrametric} structure,
meaning that their mutual distance can be described through a taxonomic or genealogical tree~\cite{mezard:84,mezard:85,rammal:86}.

% In the replica formalism, each replica is equivalent to all the others and all the mutual information about a pair of equilibrium configurations is encoded in the overlap, measuring their mutual distance. These two features are based on \emph{stochastic stability} and \emph{ultrametricity} concepts~\cite{ghirlanda:98, parisi:00origin} according to which the distance between the different states is such that they
% can be put on a taxonomic or genealogical tree. The distance between such states is compatible with their specific position on the tree.

\paragraph{The replica structure of two common MF spin glass models.} 
For the sake of concreteness, let's consider two paradigmatic MF spin glass models. Both have a low-temperature SG phase, but with different levels of RSB, and therefore present different free energy minima structures.
\begin{itemize}
\item The \textbf{spherical $\boldsymbol{p}$-spin} model has Hamiltonian 
\begin{equation}
\mathcal{H}=-\sum_{i_1,...,i_p=1}^{N} J_{i_1,i_2,...,i_p} s_{i_1}...s_{i_p},
\label{p-spin}
\end{equation}
where $p$ indicates $p$-wise interactions to which each spin participates. 
Spins are continuous variables subject to the global constraint $\sum \limits_{i=1}^N s_i^2=N$, and the random couplings are extracted from a Gaussian distribution $P(J)=\exp \left(- J^2 \frac{N^{p-1}}{p!} \right)$, where the factor $N^{p-1}$ guarantees the extensivity of the free energy in the thermodynamic limit. 
For $p \ge 3$, this class of mean-field systems is characterized by a $1$-RSB low-temperature phase, with an emergent number of minima that grows exponentially with $N$, and those are separated by extensive barriers.

Small variations to this model result in significantly different physical behaviors. In particular, setting $p=2$ results in only one single free-energy minimum (so there is no SG phase)~\cite{dedominicis:06}, whereas replacing the spherical with Ising spins~\cite{derrida:80} results in the 1-RSB phase becoming unstable toward a full-RSB phase at low temperatures~\cite{deoliveira:97,crisanti:03b,crisanti:04c}.

\item The \textbf{Sherrington-Kirkpatrick} model (SK)~\cite{sherrington:75,sherrington:78} has a Hamiltonian with an external uniform field $h$ 
 \begin{equation}\label{eq:sk}
 \mathcal{H}=-\sum \limits_{(ij)} J_{ij}s_i s_j -h \sum \limits_{i}s_i
 \end{equation}
where the sum runs over distinct pairs. The spins are Ising variables, $s_i=\pm 1$, and the random couplings are extracted from a Gaussian distribution $\mathcal{P} \simeq e^{-N J_{ij}^2/(2J^2)}$ with zero mean and variance $J^2/N$. The low-temperature phase of this model is characterized by a hierarchical organization of energy minima, as given by the Parisi solution. The emergent number of minima is sub-exponential in system size, and those are separated by sub-extensive barriers~\cite{mezard1984_SG}. The transition from single equilibrium (RS) to multiple equilibria (full-RSB) is second-order-like, with diverging correlation lengths and power-law singularities.
\end{itemize}

The solution of the SK model obtained by the replica method was rigorously proven thirty years later~\cite{guerra:02,guerra:03,talagrand2002high, talagrand:06}. Even though there is no rigorous demonstration that the replica method is generally correct, it has since been proven to provide the correct result in several specific cases~\cite{brennecke2022replica, shinzato2018, barbier:19}, and no counter-example is known. %of it providing wrong results.

\subsection{The Thouless-Anderson-Palmer (TAP) Approach}\label{sec:tap}

The TAP approach aims to probe complex energy and free-energy landscape properties through a perturbative high-temperature expansion (also known as \emph{Plefka} or \emph{Georges-Yedidia expansion}~\cite{plefka:82, georges:91}) by defining a Legendre transform $\mathcal{F}[\textbf{m}]$ of the free energy as a function of the average magnetization $\textbf{m}$. 
Such an expansion can detect the metastable states of the system, which correspond to the local minima of an appropriately-defined non-convex functional. 

The underlying idea is to enforce the system to have single-site magnetization $m_i$ by means of Lagrange multipliers $\lambda_i^{(\beta)}$ that depend on the inverse temperature $\beta$. Then, from the Legendre transform, one obtains
\begin{equation}
\mathcal{F}^{(\beta)}[\boldsymbol{m}]=\log \sum \limits_{\boldsymbol{s}} e^{-\beta \mathcal{H}[\boldsymbol{s}]+\sum\limits_{i}\lambda_i^{(\beta)}(s_i-m_i)}
\end{equation}
given the stationarity condition $\lambda_i^{(\beta)}=-\frac{\partial \mathcal{F}^{(\beta)}[\boldsymbol{m}]}{\partial m_i}$.
In the  $\beta \rightarrow 0$ limit, spins are uncorrelated, making the computation of the first derivatives \footnote{In the thermodynamic limit all higher-order terms but the first two can be neglected in a fully-connected system.} of the free-energy functional with respect to $\beta$ very straightforward:
% \begin{equation}
\begin{align}\label{TAP}
& \left . \frac{d \mathcal{F}^{(\beta)}[\boldsymbol{m}]}{d\beta}\right \vert_{\beta=0}= -\langle \mathcal{H}\rangle\,,\\
& \left . \frac{d^2 \mathcal{F}^{(\beta)}[\boldsymbol{m}]}{d\beta^2}\right \vert_{\beta=0}=\biggl \langle \left[\mathcal{H}[\boldsymbol{s}] -\langle \mathcal{H}\rangle -\sum \limits_i \frac{\partial \lambda_i^{(\beta)}}{\partial\beta} (s_i-m_i)\right]^2 \biggr \rangle %=\frac{1}{2}\sum \limits_{i \neq j} J_{ij}^2(1-m_i^2)(1-m_j^2) \ .
\,.
\end{align}
% \end{equation}
The second derivative involves both the connected part of the Hamiltonian and the partial derivatives of the Lagrange multipliers. 
The TAP free energy can then be written as a second-order expansion around the $\beta=0$ limit. For the SK model, it reads
\begin{equation}
\mathcal{F}_\mathrm{TAP}^\mathrm{SK}[\boldsymbol{m}]=-\frac{1}{\beta} \sum \limits_i s_0(m_i)-\frac{1}{2}\sum \limits_{i\neq j}J_{ij}m_i m_j -\frac{N \beta}{4}(1-q)^2\,,
\end{equation}
where the first term on the right-hand side accounts for entropic effects, the second term is average energy contribution, and the last piece, the \emph{Onsager reaction term}, represents a first correction to the MF approximation.
It is also worth noting that since the couplings in the SK model are of order $1/\sqrt{N}$, $O(\beta^2)$ terms do matter in the final TAP expression. This contrasts with the purely ferromagnetic case (the fully-connected Ising model), for which the couplings are of order $1/N$.

In the case of the spherical $p$-spin model, the TAP free energy reads
\begin{equation}
\mathcal{F}_\mathrm{TAP}^\mathrm{pspin}[\textbf{m}]=-\frac{1}{\beta} \sum \limits_{i}s_0(m_i)-\frac{1}{p!}\sum \limits_{i_1,...,i_p}J_{i_1,...,i_p}m_1...m_p -\frac{\beta N}{4} \left[ 1-pq^{p-1}+q^{p}(p-1)\right]\,,
\end{equation}
where, again, the last term represents the Onsager reaction term. By setting $m_i=0$, one can immediately check that the TAP free energy is equal to $-\beta/4$, which is the free energy of the paramagnetic state.

% Dynamically, ergodicity is broken at a given temperature, which is higher than the static temperature, because of the presence of many metastable states.
% To properly describe the emergence of even more complex energy landscapes, we shall focus on the following lines on the spherical $p$-spin model. Furthermore, its thermodynamic emerging properties have been shown to be intrinsically linked to the phenomenology of structural glasses \cite{crisanti:92, crisanti:95} and the representative dynamical equations are well captured by the same functional form as Mode Coupling Theory \cite{reichman:05}.

In principle, the equilibrium probability distribution -- hence the partition function -- can always be decomposed into a combination of \emph{pure states} $\alpha$, with a free-energy density $f_\alpha$. These pure states are entirely determined by a set of local observables, such as local magnetizations. The partition function can therefore be expressed as a sum over such states
\begin{equation}
Z= e^{-\beta N F} \simeq \sum \limits_{\alpha} e^{-\beta N f_\alpha}= \int df \; \sum \limits_{\alpha} \delta(f-f_{\alpha}) e^{-\beta N f}= \int df \; e^{N\left[ \Sigma(f,\beta)-\beta f \right]} \simeq
e^{N\left[ \Sigma(f^{*},\beta)-\beta f^{*}\right]}\,,
\end{equation}
where in the last step we applied the saddle-point method~\cite{withamlinear,bernardini1999metodi}.
The configurational entropy $\Sigma$ represents the logarithm of the total number of states. Evaluated at the states with $f^*$, which contribute maximally to $Z$, it leads to a simple temperature dependence of the TAP free energy with the identification $\partial \Sigma(f,\beta)/\partial f\vert_{f^{*}(\beta)}=\beta$. 

In the $p$-spin model, this decomposition in pure states discloses a rich behavior in terms of the number of pure states that, at a given temperature, participate in the thermodynamic behavior of the system. 
At high temperatures, the free-energy density is ruled by
a single state, the paramagnetic (Boltzmann-Gibbs) state.
Upon lowering the temperature, one reaches a \emph{dynamical transition} at temperature $T_d$, at which the emergence of an exponentially large number of clusters of pure states is accompanied by a dramatic slowing down of the dynamics. Surprisingly, the transition is not associated with any thermodynamic transition, because the free energy preserves its analyticity. The thermodynamic transition occurs at a lower temperature, $T_s$, known as \emph{static} or \emph{condensation transition}, where all different clusters of states collapse to the same state as a clear signature of vanishing configurational entropy \cite{crisanti2003com}.

\bigskip
Although the TAP approach was conceived as a high-temperature expansion, it can also be developed in terms of other perturbative quantities, such as a fictitious coupling associated with an effective energy cost. In this case, the Plefka-like expansion offers a playground for the definition of suitable effective high-dimensional potentials in situations where the energy is either ill-defined or trivially zero, such as in non-convex constraint satisfiability problems~\cite{altieri2016jamming, altieri2018}.

\subsection{The Cavity Method}
{The cavity method~\footnote{In computer science and artificial intelligence, the cavity method is also known as \emph{{Belief Propagation}}, whereas in statistical physics is referred to as \emph{Bethe-Peierls}~\cite{mezard2009information}.} was initially devised as a way to solve the SK model without needing to resort to the replica formalism, but can also be used to go beyond the MF approximation. It can notably consider correlations between spins,
% {\color{red} Anyway, we will introduce it here as an attempt to go beyond the MF approximation and consider correlations between spins.
% }
thanks to the fact that it is exact in systems with a loopless interaction graph -- or with divergingly wide loops -- such as in trees and some random graphs.
In the following, we will focus on finite-connectivity graphs, $\mathcal{G}=(V,E)$, defined by $V$ nodes and $E$ edges. 
The Hamiltonian in Eq.~\eqref{eq:sg} can be then rewritten as
$E(\textbf{s})=-\sum_{(i,j) \in E} J_{ij} s_i s_j\,,$
where $i,j$ belong to the same edge $E$. 
The graph $\mathcal{G}$ together with the set of random couplings $\Jij$ then define a sample.

The main goal of the cavity method is to calculate the probability distribution of each spin, $P(s_j)$. The idea behind the approach is that $P(s_j)$ is determined by the local fields induced by all the neighbors of $s_j$, which are in turn determined by their own neighbors, and so on (Fig.~\ref{factorgraph}). 
\begin{figure}[h]
\centering
\includegraphics[scale=0.77]{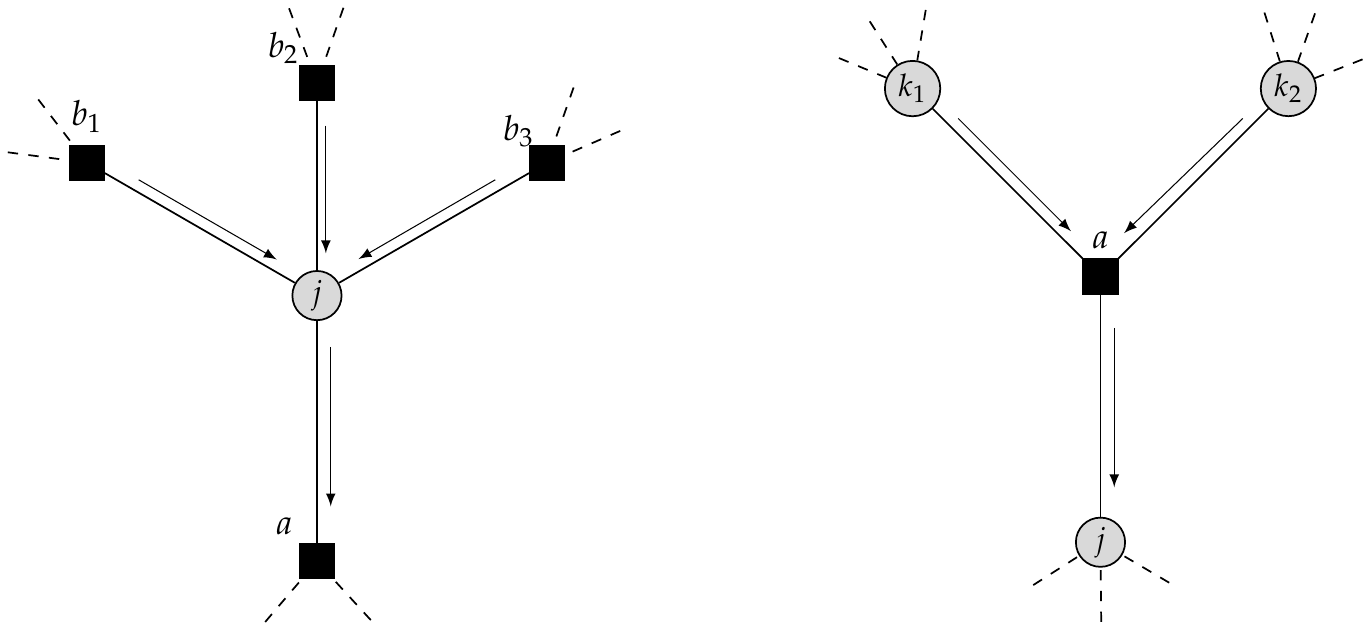}
\caption{Left: Graph representation for the message $\hat{\nu}_{b \rightarrow j}(s_j)$, from the function node (square) to the variable node (circle). Right: corresponding part of the graph for the computation of the message $\hat{\nu}_{a \rightarrow j}(s_j)$, function of $\nu_{k \rightarrow a} (\sigma_k)$ from the variable node to the function node.}
\label{factorgraph}
\end{figure}
Because of the tree-like structure of the graph, the neighbors of $s_j$ are mutually independent, and hence $P(s_j)$ can be factorized and determined through  the incoming local fields, $\hat{\nu}_{a \rightarrow j}(s_j)$ (where $a$ indicates a neighboring interaction):
$P(s_j) \simeq \prod_{a \in \partial j} \hat{\nu}_{a \rightarrow j}(s_j)$.
At the same time, the interaction $a$ influencing $s_j$ is defined by the marginal distribution of all the spins $k$ in the neighborhood of $a$, once $s_j$ is removed, $\nu_{k\to a}(s_j)$. 
Therefore, the cavity formalism results in a set of coupled equations for the marginal probability laws 
\begin{equation}
\begin{cases}
 & \nu_{j \rightarrow a}(s_i) \propto \prod \limits_{b \in \partial j \backslash a } \hat{\nu}_{b \rightarrow j}(s_j) \ ,\\
   & \hat{\nu}_{a \rightarrow j} (s_j) \propto \sum \limits_{\textbf{s} \in \partial a \backslash j} \psi_a(\textbf{s}_{\partial a}) \prod \limits_{k \in \partial a \backslash j }\nu_{k \rightarrow a}(s_k) \,,
\end{cases}
\end{equation}
where $\psi_a$ is called \textit{compatibility function}, which is analogous to the Boltzmann weight, and $\partial a \backslash j$ indicates the spins near $a$ excluding $j$.

% These marginal probabilities $\hat{\nu}_{a \rightarrow j}$ and $\nu_{j \rightarrow a}$ basically tie the function node to the variable node and vice versa. 
% By gathering all incoming messages, the single-variable marginal distribution corresponds to $P(s_i) \simeq \prod_{a \in \partial i} \hat{\nu}_{a \rightarrow i}(s_i)$.

% In the simplest scenario of Ising spins, these messages can be re-expressed in terms of the so-called \emph{cavity fields}, which precisely act as external fields once a variable is removed from the network.

The resulting free energy as a function of fixed-point messages turns out to be a combination of three contributions coming from: all function nodes, all variable nodes, and the edges connecting $i$ to any possible function node (see Fig.~\ref{factorgraph}).
When the size of the loops is larger than the correlation length, the signals entering $j$ are independent. In this case, if there is a unique solution for the state in $j$, the cavity method can recover it exactly. For instance, at the RS level, where there is only one pure state, the cavity 
approach is formally equivalent to the replica method~\cite{mezard:87}. 
% More precisely, when cutting the links between variables in $\partial j \backslash a$ and $j$, the resulting variables in the modified system are far away from each other and this implies a fast decay of the correlation with distance, provided that a unique state exists. 
%Even though the (RS) cavity equations no longer apply strictly speaking in the spin-glass phase and in the presence of short loops, they still provide a good approximation and first estimate for thermodynamic quantities of interest.
It the system is not RS, or if the loops are short, the approach needs to be refined. For example,
% To cope with the fact that (RS) cavity equations no longer apply strictly speaking in the spin-glass phase and in the presence of short loops,
$1$-RSB cavity protocols have been developed to extract properties of the pure state decomposition and satisfiability conditions in generic optimization problems \cite{braunstein:04, mertens:06}. Such $1$-RSB message passing equations go by the name of 
\emph{survey propagation} and have been generalized as \emph{population dynamics algorithms}.

%\begin{equation}
%\mathcal{F}^{*}(\boldsymbol{\nu})=\sum_{\delta \in F} \mathcal{F}_{\delta }(\boldsymbol{\nu})+ \sum_{i \in V} \mathcal{F}_i(\boldsymbol{\nu})-\sum_{(i, \delta) \in E}\mathcal{F}_{i \delta}(\boldsymbol{\nu})
%\end{equation}

\paragraph{Beyond Mean-Field: multi-layer construction and loop corrections}

The cavity method stops being exact in a non-tree-like topology and is indeed hindered by the presence of loops. Furthermore, because of its non-perturbative nature, any small parameter expansion used to compute corrections to mean-field theory \cite{efetov1990, parisi2006loop} appears to be unfeasible. One can nevertheless build a $M$-layer model, where $M$ copies of the original lattice are stacked on top of each other assuming the same distribution of random couplings. The idea -- originally proposed in computer science by Vontobel \cite{vontobel2013} -- has been reworked recently in disordered systems based on a rewiring procedure. By inducing random permutations of the links, inter-layer connections, and hence spatial loops, are automatically generated by a generalized transfer matrix formalism on uncorrelated one-dimensional chains.
The perturbative computation of finite-size corrections in powers of $1/M$ notably offers a reliable formal method to study critical phenomena in finite-dimensional systems~\cite{altieri2017loop}. %A special emphasis is then devoted to regions close to the critical point, where deviations from mean-field theory are expected, and even stronger from fully-connected models. 
The advantage of this tree-based approach is that it can recover phase transitions that deviate from the behavior of fully-connected models and are instead more similar to the finite-dimension phenomenology.
This situation arises, for instance, in the Random Field Ising Model~\cite{biroli:18_random, angelini2020loop}, which is strongly affected by non-perturbative effects in low dimensions, and in Anderson localization~\cite{abou:73} in the quantum realm.

%Nevertheless, given the large degeneracy of TAP minima, correctly weighting them  remains a very delicate aspect. The problem was partially solved in the eighties by De Dominicis and Young \cite{dedominicis} and further investigated in the years aiming to prove the equivalence between the TAP approach \cite{crisanti:95} -- which will be properly introduced in the next Section -- and standard thermodynamic methods as the replica trick. The use of the BRST symmetry \cite{aspelmeier:04, crisanti2004spin} is one of these.

\subsection{Dynamical Mean-Field Theory (DMFT)}
\label{dynamics}

Understanding the dynamics of SGs is one of the oldest and most challenging problems in the theory of matter. {One characteristic feature of the SG dynamics is that its relaxation depends on the history of the system itself, \textit{i.e.,} it ages. In other words, the autocorrelation between a time $t$ and $t'>t$ does not only depend on $(t'-t)$, but also on the \textit{age} of the system, $t$. 
{%\color{red} 
An aging system relaxes extremely slowly without ever reaching an equilibrium state. During its exploration, the system wanders across states that are not the relevant ones from a thermodynamic viewpoint and whose static properties cannot be defined rigorously.}

%In many contexts, one needs to optimize a very rough energy function in order to find, in the best possible scenario, the ground state or, less optimistically, local minima that can trap the dynamics. 
A MF solution for the dynamics in glassy systems was first suggested by Sompolinsky and Zippelius for equilibrium properties~\cite{sompolinsky:82}. Intriguing off-equilibrium features came emerged from a deeper investigation of specific models, for which a closed set of integro-differential equations could be solved~\cite{cugliandolo:93, cugliandolo1994_aging,franz1999}.

For the sake of simplicity, in the following, we shall consider the over-damped Langevin dynamics of a system in contact with a thermal bath,
\begin{equation}
 \frac{ds_i}{dt}=-\frac{\partial \mathcal{H}}{\partial s_i} +\eta_i(t)  
\end{equation}
whose behavior is captured by Gaussian white noise with zero mean and variance $\langle \eta_i(t)\eta_j(t')\rangle=2 T \delta_{ij} \delta(t-t')$.
The index $i=1,...,N$ runs over the total number of spins in the system.
The Hamiltonian $\mathcal{H}$ typically incorporates a single-spin potential $V(s)$ plus a disordered part, which explicitly depends on the quenched disordered couplings (see \textit{e.g.} Eq.~\eqref{eq:sk}), which need to be averaged out. The dynamical MF procedure allows to average over the couplings and coarsen the time-dependence of the system as that of a single average spin $s(t)$.
For the $p$-spin model, the resulting DMFT equation reads~\cite{sompolinsky:82,castellani:05}:
\begin{equation}\label{eq:dmft}
\dot{s}(t)=-\frac{\partial V(s(t))}{\partial s} + \frac{p(p-1)}{2}\int_{0}^{t} dt{'} \; R(t,t') C^{p-2}(t,t') s(t') +\xi(t)
\end{equation}
where the noise $\xi(t)$ accounts for the interaction with the rest of the system and has an extra colored term, \emph{i.e.} $\langle \xi(t) \xi(t')\rangle=2T \delta(t-t')+C(t,t')$.

The dynamical equation above is expressed in terms of the two-time correlation function 
\begin{equation}\label{eq:c}
C(t,t')=\frac{1}{N}\sum \limits_i s_i(t)s_i(t') \ ,
\end{equation}
and of the response function to an external pulse perturbation
\begin{equation}\label{eq:r}
R(t,t')=\left . \frac{1}{N}\sum \limits_i \frac{\delta  s_i(t)}{\delta h_i(t')}\right \vert_{h_i=0} \ .
\end{equation}
In the thermodynamic limit, the above equations converge to a unique non-fluctuating solution, which is the only one allowed by causality.
At equilibrium, $C(t-t')$ and $R(t-t')$ are related by the fluctuation-dissipation theorem, $R(t-t')= -\frac{1}{T} \Theta(t-t') \frac{d}{dt} C(t-t')$.
While, for finite $N$, the correlation and response functions are expected to decay exponentially in time, this property is no longer generically true in the thermodynamic limit, as the relaxation time might diverge, thus signaling a dynamical transition (Sec.~\ref{sec:tap}).

Based on the long-time limit analysis first performed in the spherical $p$-spin model~\cite{cugliandolo:93}, one can separately analyze the fast regime, in which relevant degrees of freedom rapidly equilibrate preserving time translation invariant (TTI) properties, and the slow regime, in which violations of fluctuation-dissipation relations and non-equilibrium phenomena emerge.
In the large-time limit, for $t,t'\rightarrow \infty$, the correlation function can be split as $C(t,t')=C_{\text{TTI}}(t-t')+C_\text{Aging}(t,t')$. 
Depending on the appearance either of a single diverging timescale or a multiplicity of progressively slower timescales, the slow part of the correlator, $C_\text{Aging}(t,t')$, can be captured by a combination of involved functions each associated with a slow timescale~\cite{bouchaud:98}.

Notably, by assuming the MF picture of aging as a starting point (such as Eq.~\eqref{eq:dmft}), it is possible to investigate the properties of asymptotic regimes without explicitly solving the integro-differential equations for the correlation and response functions (\textit{e.g.} Eq.~\eqref{eq:c}, \eqref{eq:r}). 
The outcome is different depending on whether the low-$T$ phase is $1$-RSB or full-RSB~\cite{altieri2020dynamical}. For the latter, a dynamical effective stochastic process for the slow-evolving effective part of Eq.~\eqref{eq:dmft} has been worked out\cite{altieri2020dynamical}, which exactly maps into the ultrametricity condition associated with the Parisi solution in the replica formalism~\cite{mezard:87}.
Hence, DMFT both in the presence of a single slow timescale and in the ultrametric scenario explicitly links the aging dynamics with the replica-based description.
%More precisely:
%\begin{equation}
%\frac{d h(x)}{dx}=\frac{x}{T} m(h) \dot{q}(x)+\sqrt{\dot{q}%(x)} z(x) \ ,
%\end{equation}
 %where $z(x)$ represents a Gaussian white noise, $\delta$-correlated in $x,x'$, and the magnetization $m(h)$ has to be averaged over the distribution of effective fields $P(h,x)$. The dynamical evolution is thus measured in terms of the effective temperature change $x=T_\text{eff}/T$.

\section{Spin Glasses in Three Dimensions}
\subsection{Numerical Methods}
Although MF treatments provide elegant exact solutions for SGs defined on a fully-connected or tree-like graph, extending those findings to three dimensions is quite challenging.
For instance, even the existence of a SG phase in the $3d$ EAM has yet to be proven analytically. Most progress, therefore, arise from numerical simulations. In particular, there is convincing numerical evidence that, as their experimental counterparts, SG models in three dimensions exhibit a continuous phase transition at a temperature $T_c$~\cite{ogielski:85b,palassini:99,ballesteros:00,mari:02,nakamura:03,daboul:04,pleimling:05,perez-gaviro:06,parisen:06,jorg:06,campbell:06,katzgraber:06,machta:08,hasenbusch:08,hasenbusch:08b,fernandez:09b,janus:13}. 

The low-temperature dynamics of SGs is, however, very slow, due to the competition between short and long-range interactions, and the presence of temperature chaos. In the SG phase, the free energy profile indeed changes drastically even for infinitesimal changes in temperature~\cite{seoane:13}. The numerical study of SGs at low $T$ is therefore highly challenging~\cite{yllanes:11}. Numerical strategies advanced for studying SGs notably include: the construction and use of special-purpose computers~\cite{pearson:81,condon:85,vandersijs:96,cruz:01,janus:06,janus:08,janus:12b,janus:13c,janus:14}, GPUs~\cite{weigel:11,weigel:12,fang:14,baityjesi:14,lulli:15}, the formulation parallelization techniques, such as multi-spin coding~\cite{newman:99,fang:14,lulli:15,kawashima:96,isakov:15,jorg:08b,fernandez:16b}, which, by encoding every spin on a single bit and restricting to bit-to-bit operations, allows one to simulate $64$ replicas (the number of bits in a long integer) in the time of one;
% On another hand, several methods were developed for the simulation of SGs. One example is \textit{quiet planting}, which for creating an equilibrium configuration at $T$, first generates the spin configuration $\{s_i\}$ and then finds the couplings such that  $\{s_i\}$ is precisely at equilibrium ~\cite{ozeki:97,montanari:06b,krzakala:09,krzakala:11,krzakala:11b}. 
and the development of algorithms, such as \textit{parallel tempering}, which consists of simultaneously simulating several replicas at different temperatures, and proposing Monte-Carlo updates, which exchange the temperatures among replicas~\cite{swendsen:86,hukushima:96,wang:15b}.

\subsection{Spatial Correlations}
A key feature of low-dimensional systems is the characteristic extent of spatial correlations, $\xi$. 
Its typical experimental determination is through the magnetic response to an external magnetic field $h$~\cite{joh:99,nakamae:12,guchhait:14,guchhait:17}, which provides the coherence length $\xi_\mathrm{Zeeman}$.
In numerical simulations, the size of the correlated domains is instead calculated through correlation functions of the form~\cite{janus:10,janus:09b}
\begin{equation}
    C(\rr) =\frac1N \sum_{\x}^N q_{\x}q_{\x+\rr}\,,
\end{equation}
where $\x$ indicates a position in the lattice, $\rr$ is a displacement vector, and $q_\x$ is usually the overlap at site $\x$, $q_\x=q^{(ab)}_\x=s_\x^{(a)}s_\x^{(b)}$, or the link-overlap, $q_\x=q^{(ab)}_{\mathrm{link,\x}}=q_\x^{(ab)}q_{\x+\hat{e}}^{(ab)}$ ($\hat{e}$ is a unit vector)~\cite{caracciolo:90,altieri2016_RG}, since these quantities are equivalent in the limit $d\rightarrow\infty$. There are several ways to extract a correlation length $\xi_\mathrm{micro}$ from $C(\rr)$~\cite{cooper:82,janus:08b,janus:09b,baityjesi:16}. For example, $\xi_\mathrm{micro}=\sqrt{\frac{\int r^2 C(r) dr}{\int C(r)}}$. While, out of equilibrium, $\xi_\mathrm{Zeeman}$ and $\xi_\mathrm{micro}$ have different behaviors~\cite{janus:23}, at equilibrium they match~\cite{janus:17b}.
These length scales can be used to identify phase transitions through finite-size scaling~\cite{fisher:72,nightingale:76,binder:81}, by comparing simulations performed in systems of different linear sizes $L$. Because quantities such as $\frac{\langle\xi(T)\rangle}{L}$ are scale-invariant at $T_c$, one can identify critical points by investigating where the curves $\frac{\langle\xi(T)\rangle}{L}$ cross for different $L$. %In addition to $\frac{\langle\xi(T)\rangle}{L}$, 
For the study of SGs, it has also proven useful to consider other quantities, such as ratios of susceptibilities~\cite{janus:12}, possibly conditioned to given values of the overlap~\cite{janus:14c}.

\subsection{The Spin-Glass Phase in Low Dimensions}
\paragraph{Pictures for the nature of the low-temperature phase in three dimensions.}
Just like there is no rigorous proof of the existence of $T_c$ in three dimensions, there is no first-principles theory on the nature of the spin-glass phase in this case either\footnote{
% {\color{red}There is no clear evidence on the nature of the spin-glass phase in 3$d$, although pioneering studies on finite-dimensional interfaces have shown interesting behavior even in this case}.
The results in this section refer to the $3d$ EAM, which is the most studied model.}. Two main phenomenological scenarios have been proposed: the Droplet and the RSB pictures. The former~\cite{fisher:87,huse:87,fisher:88,fisher:88b} is based on a renormalization group approach on the EAM~\cite{migdal:75,kadanoff:76}, which is exact in one dimension, and depicts the SG phase as having only two pure states, with $q=\pm\qea$, reminiscently of the behavior of the ferromagnetic Ising model~\cite{huang:87}. One consequence of this proposal is that the size of the surface separating different magnetic domains scales as $L^{d_s}$, with $d_s<d$ (for the Ising ferromagnet $d_s=d-1$), and the energy of the smallest excitation grows with $L$. Another consequence is that the low-temperature phase disappears as soon as an external magnetic field is applied to the system.\\
The RSB picture~\cite{parisi:96,marinari:00,parisi:12} is based on the opposite limit, $d=\infty$. It assumes that the SG phase of the EAM is qualitatively similar to that of the SK model, with an overlap distribution $P(q)$ which is non-zero over a wide interval of $q$. This picture prescribes that the domain surfaces are space-filling ($d_s=d$) and the smallest excitation remains $\mathcal{O}(1)$ when $L\to\infty$. In addition, the SG phase survives when a finite magnetic field is applied, with a dAT line $h_c(T)$ separating the paramagnetic from the SG phase (as described in Sec.~\ref{sec:replica}). Whether either of the two theories holds remains, however, a matter of debate~\cite{moore:11,parisi:12,yeo:12,yucesoy:12,yucesoy:13,billoire:13,ruizlorenzo:20,holler:20,moore:21,newman:22,martin-mayor:22}.

\paragraph{Evidence on the nature of the spin glass phase in $3d$.}
% https://arxiv.org/search/?query=%22janus+collaboration%22&searchtype=all&abstracts=show&order=-announced_date_first&size=50

% Given the vague-definedness of the pure states in a spin glass~\cite{metastates}, and how hard it is to identify them numerically~\cite{metastates-identification}, most of the numerical research on the nature of the SG phase consisted of searching quantitative confirmations/falsifications of the Droplet and RSB scenarios.

Evidence to falsify or support the Droplet and RSB scenarios has principally been sought through numerical simulations. 
For example,
equilibrium runs of the $3d$ EAM with linear size $L\leq32$ deep in the SG phase show that $P(q)$ has wide support, and $P(0)$ is stable as $L$ increases~\cite{janus:10,janus:11}. This observation is in contradiction with the Droplet prediction, $P(q)\propto[\delta(q-\qea)+\delta(q+\qea)]$, which assumes $P(0)=0$.\\
When an external magnetic field is applied to the system, equilibrium simulations appear perturbed by finite-size effects~\cite{janus:14c}, and there is disagreement on the existence of a dAT line. %, because overlap distributions are fat-tailed~\cite{monthus:13,janus:14c}. 
Arguments stemming from numerical simulations on the $3d$ EAM state that a phase transition could be present at temperatures that are 40\% lower than those accessible~\cite{janus:14b,janus:14c}. Those are opposed by arguments from simulations on SG chains with long-ranged interactions, which mimic $3d$ systems, claiming that no transition exists~\cite{vedula2023}.
% Numerical simulations are currently unable to detect a hypothetical transition because it is thought to lay at a lower $T$ than those accessible (provided that it exists)~\cite{janus:14b,janus:14c}. 
We note, however, that the upper critical dimension is $d_\mathrm{up}\geq6$~\cite{mezard:87,angelini:22}, and that there is strong evidence for a dAT transition in $d=4$~\cite{janus:12}. 
Perturbative renormalization group analysis in $d=6-\epsilon$ is consistent with the presence of a dAT line~\cite{temesvari:23,parisi:12,temesvari:08,holler:20}, with a non-trivial fixed point appearing at second order in $\epsilon$~\cite{charbonneau:17b,charbonneau:19}.

% The size of the excitations was also measured, and both numerical~\cite{krzakala:00,maiorano:18} and analytical evidence~\cite{franz:92,franz:94,astuti:19} exclude that they are extensive.

% {Fractal dimension} - 
% RSB: \cite{}
% Droplet: \cite{wang:18}

% \subsection{Calculating the lower critical dimension}
% \subparagraph{Metastable states} Newman and Read
% \cite{franz:95b}

\paragraph{Outlook on the nature of the spin glass phase}
The current numerical evidence does not confirm the Droplet picture, but it has been argued that the observations might change if larger system sizes were simulated~\cite{moore:21, vedula2023}.
It is to be noted, however, that times and system sizes of numerical simulations nowadays closely approach those of experiments~\cite{janus:10,janus:18,zhai:19,paga:22}. Therefore, if even there is a limit of very large $L$ in which the Droplet theory applies, this might not be experimentally relevant~\cite{martin-mayor:22}.
Evidence in favor of the RSB picture is, however, not decisive, and relevant beyond-MF mechanisms might arise in low dimensions.
Alternative or intermediate scenarios could also be valid~\cite{holler:20}, such as the trivial-non-trivial picture~\cite{palassini:00,krzakala:00}, according to which the domain surfaces are not space-filling, as in the Droplet picture, but there are large-scale excitations whose energy does not increase with size~\cite{palassini:03,chatterjee:23}, as prescribed by the RSB scenario. 

\section{Conclusions}

\paragraph{Universality classes}
While our discussion thus far focused on few specific SG models, changing details of the Hamiltonian can change general features of the energy landscape, such that new physics emerges.
As discussed above, dimensionality plays a crucial role, with the $d=1,\infty$ limits having dissimilar behaviors that might both differ from $d=3$. Altering the distribution of the couplings~\cite{cizeau:93,janzen:10}, $P(J_{ij})$, does not seem to influence the universality class of the models as long as they are not fat-tailed~\cite{hasenbusch:08,andresen:11,janzen:10}.
Changing the interaction \textit{range} is also a relevant perturbation, in that a system with longer-ranged interactions is more MF like~\cite{katzgraber:03}. This effect has notably led to the search for a correspondence between short-ranged high-dimensional models and long-ranged SGs in $d=1$~\cite{katzgraber:05,banos:12,leuzzi:13,wittmann:16,vedula2023}.

Another important is the nature of the spins.
As described in Sec.~\ref{sec:replica}, one can soften spins, passing from Ising to spherical spins. For pairwise interactions in $d=\infty$, this completely changes the landscape, which passes from complex to simple. (The dynamics nevertheless remains slow due to a large number of flat directions~\cite{cugliandolo:93,cugliandolo:94, barrat:98}.)
Alternatively, one can use spins that are normalized vectors with $m$ components~\cite{dealmeida:78b}. For high $m$, the landscape becomes trivial, both in high-~\cite{hastings:00} and low-dimensional systems~\cite{baityjesi:15}, which points toward using the limit $m\to\infty$ for studies of the $1/m$ expansions~\cite{aspelmeier:04b,moore:12}. Because the $m\to\infty$ behavior seems radically different from any finite $m$~\cite{lee:05}, however, this approach has not been extensively pursued.

One can also consider Hamiltonians that mix interactions with different numbers of bodies, such as \emph{mixed $p$-spin models}~\cite{barrat:97,crisanti:04}. These models display several crossover temperatures at which the dynamical behavior changes qualitatively, apparently without any thermodynamic transitions~\cite{folena:20,folena:21,folena:20b}. 
This behavior is reminiscent of that of supercooled liquids~\cite{debenedetti:21}.

% seem to resemble the phenomenology in $3d$ structural glasses, with several crossover temperatures where the dynamical behavior qualitatively changes, apparently without transitions in the equilibrium behavior~\cite{folena:20,folena:21,folena:20b}. 
% Relaxing the system from finite temperature and looking at long-time relaxation dynamics, an exponential number of marginally stable states appear over a finite energy range, without being no longer the threshold states that we have presented for the pure $p$-spin.

An important aspect, which goes beyond the description of the specific model, is that over the years SGs have become a theoretical paradigm for modeling many different complex systems across disciplines, with an extremely broad spectrum of equilibrium and out-of-equilibrium properties. 
%Nowadays, we find spin glasses in the domain of ecology~\cite{biroli:18,roy:19,roy:20,altieri:21, Altieri2022}, finance~\cite{galluccio:98}, machine learning~\cite{choromanska:15,kawaguchi:16,baldassi:16,baityjesi:18b2}, signal reconstruction~\cite{ros:19,ros:19b}, optimization ~\cite{}, constraint satisfaction~\cite{mezard:02, franz2015, altieri2016jamming, franz2017, altieri2018}, random lasers~\cite{angelani:06,antenucci:15} and more, applying the same techniques mentioned thus far.
In fact, theories based on disordered systems often reveal to be beneficial for the treatment of problems outside the domain of SGs. Heterogeneous systems, either with quenched or self-generated disorder, are much more general than it might be believed at first glance, ranging from neural networks and optimization problems ~\cite{gardner:89, mezard:02, baldassi:16, mannelli2020, chaudhari:15, mertens:06, franz2015, antenucci2019, altieri2016jamming, altieri2018}, signal reconstruction~\cite{ros:19,ros:19b}, random lasers~\cite{angelani:06,antenucci:15}, financial markets~\cite{galluccio:98, garnier:2021, dessertaine:22}, supercooled liquids and jammed packings~\cite{charbonneau:14, rainone:15, altieri2018microscopic,altieri2019jamming, parisi2020} and theoretical ecology~\cite{bunin2017ecological, biroli:18, altieri2019, altieri:21, Altieri2022, roy:20, lorenzana:22, altieri2022glassy, ros:22}, and thermodynamic and dynamic formalisms rooted in spin-glass theory can still be used to solve them.  
% On the other hand, a major implication of their broad applicability also reflects on the nature of emergent collective phases. Regardless of the microscopic details of the system under investigation, which may be defined in terms of any other variable besides magnetic one, and which might be either discrete or continuous, vector-valued or scalar, the resulting phase transitions appear to be in line with pictures and models described above. 

\paragraph{Open problems}
SGs can be defined simply, but many questions remain open. We note: the nature of the SG phase \textit{[does replica symmetry breaking occur in finite-dimensional systems~\cite{moore:21,martin-mayor:22}?]}; 
their out-of-equilibrium behavior \textit{[how and which length scales describe their evolution~\cite{janus:19,zhai:19,janus:23,martin-mayor:22}? how are equilibrium states related to the closest local minima~\cite{folena:20b}?]};
or activated dynamics \textit{[how does the dynamics take place at times $t\gg N$~\cite{benarous:02,cammarota:15,gayrard:16,baityjesi:18,hartarsky:19,gayrard:19,stariolo:19,stariolo:20,ros:20,carbone:20,ros:21,rizzo:21,carbone:22}?]}; 
temperature chaos \textit{[how and why does the landscape change drastically even with small temperature changes and how can we harness it to obtain low-temperature configurations~\cite{fernandez:13,billoire:14,guchhait:15b,fernandez:16,janus:21b}?]};
the identification of metastates \textit{[can we explicitly measure the pure states of the SG phase~\cite{aizenman:90,newman:92,billoire:17,newman:23}?]};
connection to other kinds of systems \textit{[can we map the SG behavior onto other systems such as supercooled liquids or deep neural networks?~\cite{kirkpatrick:78,tarzia:07,charbonneau:14,baityjesi:15b,choromanska:15,kawaguchi:16,rizzo:16,baityjesi:18b2, baityjesi:19b,franz:19,bahri:20,folena:21}?]}.

These questions are not mere academic curiosities about a disordered magnetic alloy, because -- as it has happened several times in the past already -- spin-glass theory can significantly impact other fields. 
We take theoretical ecology as a final example. 
It is well known that the kind of sparsity of the interaction matrix of couplings between species has a crucial influence on the behavior of the ecosystem~\cite{marcus:22} and that these couplings are typically sparse in nature. Therefore, if we could harness low-dimensional SGs, which \eg have spatial fluctuations, our understanding of ecological and biological communities would significantly advance. 
%Thus, advancing SG theory will likely have a strong impact on wider scientific knowledge.

\section*{Acknowledgements}
We thank Patrick Charbonneau, Silvio Franz, Victor Martín-Mayor, Michael A. Moore, and Francesco Zamponi for careful reading and suggestions on the manuscript.

\bibliographystyle{unsrt}
\bibliography{biblio}

\begin{thebibliography}{100}

\bibitem{mydosh:93}
J.~A. Mydosh.
\newblock {\em Spin Glasses: an Experimental Introduction}.
\newblock Taylor and Francis, London, 1993.

\bibitem{binder:86}
K.~Binder and A.~P. Young.
\newblock Spin glasses: Experimental facts, theoretical concepts, and open
  questions.
\newblock {\em Rev. Mod. Phys.}, 58:801--976, Oct 1986.

\bibitem{nordblad:16}
Per Nordblad.
\newblock Disordered magnetic systems.
\newblock {\em Reference Module in Materials Science and Materials
  Engineering}, 2016.

\bibitem{Vincent:22}
Eric Vincent.
\newblock Spin glass experiments.
\newblock {\em arXiv preprint arXiv:2208.00981}, 2022.

\bibitem{ruderman:54}
M.A. Ruderman and C.~Kittel.
\newblock Indirect exchange coupling of nuclear magnetic moments by conduction
  electrons.
\newblock {\em Phys. Rev.}, 96:99, 1954.

\bibitem{kasuya:56}
T.~Kasuya.
\newblock A theory of metallic ferro- and antiferromagnetism on zener's model.
\newblock {\em Prog. Theor. Phys.}, 16:45, 1956.

\bibitem{yosida:57}
K.~Yosida.
\newblock Magnetic properties of cu-mn alloys.
\newblock {\em Phys. Rev.}, 106:893--898, Jun 1957.

\bibitem{anderson:70}
P.W. Anderson.
\newblock Localisation theory and cumn problems - spin glasses.
\newblock {\em Materials Research Bulletin}, 5:549, 1970.

\bibitem{dzyaloshinsky:58}
I.~Dzyaloshinsky.
\newblock A thermodynamic theory of “weak” ferromagnetism of
  antiferromagnetics.
\newblock {\em J. Phys. Chem. Sol.}, 4:241, 1958.

\bibitem{moriya:60}
T.~Moriya.
\newblock New mechanism of anisotropic superexchange interaction.
\newblock {\em Phys. Rev. Lett.}, 4:5, 1960.

\bibitem{fert:80}
A.~Fert and Peter~M. Levy.
\newblock Role of anisotropic exchange interactions in determining the
  properties of spin-glasses.
\newblock {\em Phys. Rev. Lett.}, 44:1538--1541, Jun 1980.

\bibitem{martin-mayor:11b}
V.~Mart\'{i}n-Mayor and S.~Perez-Gaviro.
\newblock Three-dimensional heisenberg spin glass under a weak random
  anisotropy.
\newblock {\em Phys. Rev. B}, 84:024419, Jul 2011.

\bibitem{baityjesi:14}
M.~Baity-Jesi, L.~A. Fernandez, V.~Mart\'{i}n-Mayor, and J.~M. Sanz.
\newblock Phase transition in three-dimensional heisenberg spin glasses with
  strong random anisotropies through a multi-gpu parallelization.
\newblock {\em Phys. Rev.}, 89:014202, 2014.

\bibitem{young:98}
A.~P. Young.
\newblock {\em Spin Glasses and Random Fields}.
\newblock World Scientific, Singapore, 1998.

\bibitem{guerra:02}
Francesco Guerra and Fabio~Lucio Toninelli.
\newblock The thermodynamic limit in mean field spin glass models.
\newblock {\em Communications in Mathematical Physics}, 230(1):71--79, 2002.

\bibitem{carmona:06}
Philippe Carmona and Yueyun Hu.
\newblock Universality in sherrington--kirkpatrick's spin glass model.
\newblock In {\em Annales de l'Institut Henri Poincare (B) Probability and
  Statistics}, volume~42, pages 215--222. Elsevier, 2006.

\bibitem{chatterjee:05}
Sourav Chatterjee.
\newblock A simple invariance theorem.
\newblock {\em arXiv preprint math/0508213}, 2005.

\bibitem{jagannath:22}
Aukosh Jagannath and Patrick Lopatto.
\newblock Existence of the free energy for heavy-tailed spin glasses.
\newblock 2022.

\bibitem{edwards:75}
S.~F. Edwards and P.~W. Anderson.
\newblock Theory of spin glasses.
\newblock {\em Journal of Physics F: Metal Physics}, 5:965, 1975.

\bibitem{edwards:76}
S.~F. Edwards and P.~W. Anderson.
\newblock Theory of spin glasses. ii.
\newblock {\em J. Phys. F}, 6(10):1927, 1976.

\bibitem{dealmeida:78}
J.~R.~L. de~Almeida and D.~J. Thouless.
\newblock Stability of the {S}herrington-{K}irkpatrick solution of a spin glass
  model.
\newblock {\em J. Phys. A: Math. Gen.}, 11:983, 1978.

\bibitem{bray:78}
A.~J. Bray and M.~A. Moore.
\newblock Replica-symmetry breaking in spin-glass theories.
\newblock {\em Phys. Rev. Lett.}, 41:1068--1072, Oct 1978.

\bibitem{parisi:79}
G.~Parisi.
\newblock Infinite number of order parameters for spin-glasses.
\newblock {\em Phys. Rev. Lett.}, 43:1754--1756, Dec 1979.

\bibitem{parisi:79b}
G.~Parisi.
\newblock Toward a mean field theory for spin glasses.
\newblock {\em Phys. Lett.}, 73A:203, 1979.

\bibitem{parisi:80}
G.~Parisi.
\newblock The order parameter for spin glasses: a function on the interval 0-1.
\newblock {\em J. Phys. A: Math. Gen.}, 13:1101, 1980.

\bibitem{parisi:80b}
G.~Parisi.
\newblock A sequence of approximated solutions to the s-k model for spin
  glasses.
\newblock {\em J. Phys. A: Math. Gen.}, 13:L115--L121, 1980.

\bibitem{parisi:80c}
G~Parisi.
\newblock Magnetic properties of spin glasses in a new mean field theory.
\newblock {\em J. Phys. A}, 13(5):1887, 1980.

\bibitem{mezard:87}
M.~M{\'e}zard, G.~Parisi, and M.~Virasoro.
\newblock {\em Spin-Glass Theory and Beyond}.
\newblock World Scientific, Singapore, 1987.

\bibitem{ghirlanda:98}
S.~Ghirlanda and F.~Guerra.
\newblock {\em J. Phys. A: Math. Gen.}, 31:9149, 1998.

\bibitem{parisi:00origin}
Giorgio Parisi and Federico Ricci-Tersenghi.
\newblock On the origin of ultrametricity.
\newblock {\em Journal of Physics A: Mathematical and General}, 33(1):113,
  2000.

\bibitem{mezard:84}
M.~M{\'e}zard, G.~Parisi, N.~Sourlas, G.~Toulouse, and M.A. Virasoro.
\newblock Nature of the spin-glass phase.
\newblock {\em Phys. Rev. Lett.}, 52:1156, 1984.

\bibitem{mezard:85}
M.~M{\'e}zard and M.A. Virasoro.
\newblock On the microstructure of ultrametricity.
\newblock {\em J. Physique}, 46:1293--1307, 1985.

\bibitem{rammal:86}
R.~Rammal, G.~Toulouse, and M.~A. Virasoro.
\newblock Ultrametricity for physicists.
\newblock {\em Rev. Mod. Phys.}, 58:765--788, Jul 1986.

\bibitem{dedominicis:06}
C.~de~Dominicis and I.~Giardina.
\newblock {\em {Random {F}ields and {S}pin {G}lasses}: a field theory
  approach}.
\newblock Cambridge University Press, Cambridge, England, 2006.

\bibitem{derrida:80}
B.~Derrida.
\newblock Random-energy model: Limit of a family of disordered models.
\newblock {\em Phys. Rev. Lett.}, 45:79--82, Jul 1980.

\bibitem{deoliveira:97}
Viviane~M de~Oliveira and JF~Fontanari.
\newblock Landscape statistics of the p-spin ising model.
\newblock {\em Journal of Physics A: Mathematical and General}, 30(24):8445,
  1997.

\bibitem{crisanti:03b}
A~Crisanti, L~Leuzzi, and T~Rizzo.
\newblock The complexity of the spherical $p$-spin spin glass model, revisited.
\newblock {\em The European Physical Journal B-Condensed Matter and Complex
  Systems}, 36(1):129--136, 2003.

\bibitem{crisanti:04c}
Andrea Crisanti, Luca Leuzzi, Giorgio Parisi, and Tommaso Rizzo.
\newblock Spin-glass complexity.
\newblock {\em Physical review letters}, 92(12):127203, 2004.

\bibitem{sherrington:75}
David Sherrington and Scott Kirkpatrick.
\newblock Solvable model of a spin-glass.
\newblock {\em Phys. Rev. Lett.}, 35:1792--1796, Dec 1975.

\bibitem{sherrington:78}
D~Sherrington.
\newblock Stability of the sherrington-kirkpatrick solution of a spin glass
  model: a reply.
\newblock {\em J. Phys. A}, 11(8):L185, 1978.

\bibitem{mezard1984_SG}
Marc M{\'e}zard, Giorgio Parisi, Nicolas Sourlas, G~Toulouse, and Miguel
  Virasoro.
\newblock Nature of the spin-glass phase.
\newblock {\em Physical review letters}, 52(13):1156, 1984.

\bibitem{guerra:03}
F.~Guerra.
\newblock Broken replica symmetry bounds in the mean field spin glass model.
\newblock {\em Comm. Math. Phys.}, 233:1--12, 2003.

\bibitem{talagrand2002high}
Michel Talagrand.
\newblock On the high temperature phase of the sherrington-kirkpatrick model.
\newblock {\em Annals of probability}, pages 364--381, 2002.

\bibitem{talagrand:06}
M.~Talagrand.
\newblock The {P}arisi formula.
\newblock {\em Ann. of Math.}, 163:221, 2006.

\bibitem{brennecke2022replica}
Christian Brennecke and Horng-Tzer Yau.
\newblock The replica symmetric formula for the sk model revisited.
\newblock {\em Journal of Mathematical Physics}, 63(7):073302, 2022.

\bibitem{shinzato2018}
Takashi Shinzato.
\newblock Validation of the replica trick for simple models.
\newblock {\em Journal of Statistical Mechanics: Theory and Experiment},
  2018(4):043306, 2018.

\bibitem{barbier:19}
Jean Barbier and Nicolas Macris.
\newblock The adaptive interpolation method: a simple scheme to prove replica
  formulas in bayesian inference.
\newblock {\em Probability theory and related fields}, 174(3):1133--1185, 2019.

\bibitem{plefka:82}
Timm Plefka.
\newblock Convergence condition of the tap equation for the infinite-ranged
  ising spin glass model.
\newblock {\em Journal of Physics A: Mathematical and general}, 15(6):1971,
  1982.

\bibitem{georges:91}
Antoine Georges and Jonathan~S Yedidia.
\newblock How to expand around mean-field theory using high-temperature
  expansions.
\newblock {\em Journal of Physics A: Mathematical and General}, 24(9):2173,
  1991.

\bibitem{withamlinear}
GB~Witham.
\newblock Linear and nonlinear waves, 1974.

\bibitem{bernardini1999metodi}
Carlo Bernardini, Orlando Ragnisco, and Paolo~Maria Santini.
\newblock {\em Metodi matematici della fisica}.
\newblock Carocci, 1999.

\bibitem{crisanti2003com}
Andrea Crisanti, Luca Leuzzi, and Tommaso Rizzo.
\newblock The complexity of the spherical $p$-spin spin glass model, revisited.
\newblock {\em The European Physical Journal B-Condensed Matter and Complex
  Systems}, 36(1):129--136, 2003.

\bibitem{altieri2016jamming}
Ada Altieri, Silvio Franz, and Giorgio Parisi.
\newblock The jamming transition in high dimension: an analytical study of the
  tap equations and the effective thermodynamic potential.
\newblock {\em Journal of Statistical Mechanics: Theory and Experiment},
  2016(9):093301, 2016.

\bibitem{altieri2018}
Ada Altieri.
\newblock Higher-order corrections to the effective potential close to the
  jamming transition in the perceptron model.
\newblock {\em Physical Review E}, 97(1):012103, 2018.

\bibitem{mezard2009information}
Marc Mezard and Andrea Montanari.
\newblock {\em Information, physics, and computation}.
\newblock Oxford University Press, 2009.

\bibitem{braunstein:04}
Alfredo Braunstein and Riccardo Zecchina.
\newblock Survey propagation as local equilibrium equations.
\newblock {\em Journal of Statistical Mechanics: Theory and Experiment},
  2004(06):P06007, 2004.

\bibitem{mertens:06}
Stephan Mertens, Marc M{\'e}zard, and Riccardo Zecchina.
\newblock Threshold values of random k-sat from the cavity method.
\newblock {\em Random Structures \& Algorithms}, 28(3):340--373, 2006.

\bibitem{efetov1990}
KB~Efetov.
\newblock Effective medium approximation in the localization theory: Saddle
  point in a lagrangian formulation.
\newblock {\em Physica A: Statistical Mechanics and its Applications},
  167(1):119--131, 1990.

\bibitem{parisi2006loop}
Giorgio Parisi and Franti{\v{s}}ek Slanina.
\newblock Loop expansion around the bethe--peierls approximation for lattice
  models.
\newblock {\em Journal of Statistical Mechanics: Theory and Experiment},
  2006(02):L02003, 2006.

\bibitem{vontobel2013}
Pascal~O Vontobel.
\newblock Counting in graph covers: A combinatorial characterization of the
  bethe entropy function.
\newblock {\em IEEE Transactions on Information Theory}, 59(9):6018--6048,
  2013.

\bibitem{altieri2017loop}
Ada Altieri, Maria~Chiara Angelini, Carlo Lucibello, Giorgio Parisi, Federico
  Ricci-Tersenghi, and Tommaso Rizzo.
\newblock Loop expansion around the bethe approximation through the m-layer
  construction.
\newblock {\em Journal of Statistical Mechanics: Theory and Experiment},
  2017(11):113303, 2017.

\bibitem{biroli:18_random}
Giulio Biroli, Chiara Cammarota, Gilles Tarjus, and Marco Tarzia.
\newblock Random field ising-like effective theory of the glass transition. ii.
  finite-dimensional models.
\newblock {\em Physical Review B}, 98(17):174206, 2018.

\bibitem{angelini2020loop}
Maria~Chiara Angelini, Carlo Lucibello, Giorgio Parisi, Federico
  Ricci-Tersenghi, and Tommaso Rizzo.
\newblock Loop expansion around the bethe solution for the random magnetic
  field ising ferromagnets at zero temperature.
\newblock {\em Proceedings of the National Academy of Sciences},
  117(5):2268--2274, 2020.

\bibitem{abou:73}
Ragi Abou-Chacra, DJ~Thouless, and PW~Anderson.
\newblock A selfconsistent theory of localization.
\newblock {\em Journal of Physics C: Solid State Physics}, 6(10):1734, 1973.

\bibitem{sompolinsky:82}
H.~Sompolinsky and A.~Zippelius.
\newblock Relaxational dynamics of the edwards-anderson model and the
  mean-field teheory of spin-glasses.
\newblock {\em Phys. Rev. B}, 25:6860, 1982.

\bibitem{cugliandolo:93}
L.~F. Cugliandolo and J.~Kurchan.
\newblock Analytical solution of the off-equilibrium dynamics of a long-range
  spin-glass model.
\newblock {\em Phys. Rev. Lett.}, 71:173--176, Jul 1993.

\bibitem{cugliandolo1994_aging}
LF~Cugliandolo, J~Kurchan, and F~Ritort.
\newblock Evidence of aging in spin-glass mean-field models.
\newblock {\em Physical Review B}, 49(9):6331, 1994.

\bibitem{franz1999}
Silvio Franz, Marc Mezard, Giorgio Parisi, and Luca Peliti.
\newblock The response of glassy systems to random perturbations: A bridge
  between equilibrium and off-equilibrium.
\newblock {\em Journal of statistical physics}, 97(3):459--488, 1999.

\bibitem{castellani:05}
T.~Castellani and A.~Cavagna.
\newblock Spin-glass theory for pedestrians.
\newblock {\em J. Stat. Mech.}, 2005:P05012, 2005.

\bibitem{bouchaud:98}
J.P. Bouchaud, L.C. Cugliandolo, Kurchan J., and M.~M{\'e}zard.
\newblock Out of equilibrium dynamics in spin-glasses and other glassy systems.
\newblock In A.~P. Young, editor, {\em Spin glasses and random fields}. World
  Scientific, Singapore, 1998.

\bibitem{altieri2020dynamical}
Ada Altieri, Giulio Biroli, and Chiara Cammarota.
\newblock Dynamical mean-field theory and aging dynamics.
\newblock {\em Journal of Physics A: Mathematical and Theoretical},
  53(37):375006, 2020.

\bibitem{ogielski:85b}
Andrew~T. Ogielski and Ingo Morgenstern.
\newblock Critical behavior of three-dimensional ising spin-glass model.
\newblock {\em Phys. Rev. Lett.}, 54:928--931, Mar 1985.

\bibitem{palassini:99}
M.~Palassini and S.~Caracciolo.
\newblock Universal finite-size scaling functions in the {3D} {I}sing spin
  glass.
\newblock {\em Phys. Rev. Lett.}, 82:5128--5131, 1999.

\bibitem{ballesteros:00}
H.~G. Ballesteros, A.~Cruz, L.~A. Fernandez, V.~Mart{\'i}n-Mayor, J.~Pech,
  J.~J. Ruiz-Lorenzo, A.~Tarancon, P.~Tellez, C.~L. Ullod, and C.~Ungil.
\newblock Critical behavior of the three-dimensional {I}sing spin glass.
\newblock {\em Phys. Rev. B}, 62:14237--14245, 2000.

\bibitem{mari:02}
P.~O. Mari and I.~A. Campbell.
\newblock {\em Phys. Rev. B}, 65:184409, 2002.

\bibitem{nakamura:03}
T.~Nakamura, S.-i. Endoh, and T.~Yamamoto.
\newblock {\em J. Phys. A}, 36:10895, 2003.

\bibitem{daboul:04}
D.~Daboul, I.~Chang, and A.~Aharony.
\newblock {\em Eur. Phys. J. B}, 41:231, 2004.

\bibitem{pleimling:05}
Michel Pleimling and I.~Campbell.
\newblock Dynamic critical behavior in ising spin glasses.
\newblock {\em Phys. Rev. B}, 72:184429, Nov 2005.

\bibitem{perez-gaviro:06}
S.~Perez-Gaviro, J.~J. Ruiz-Lorenzo, and A.~Taranc{\'o}n.
\newblock {\em J. Phys. A: Math. Gen.}, 39:8567--8577, 2006.

\bibitem{parisen:06}
F.~Parisen~Toldin, A.~Pelissetto, and E.~Vicari.
\newblock {\em J. Stat. Mech.: Theory Exp.}, 2006:P06002, 2006.

\bibitem{jorg:06}
T.~J{\"o}rg.
\newblock {\em Phys. Rev. B}, 73:224431, 2006.

\bibitem{campbell:06}
I.~A. Campbell, K.~Hukushima, and H.~Takayama.
\newblock {\em Phys. Rev. Lett.}, 97:117202, 2006.

\bibitem{katzgraber:06}
H.~G. Katzgraber, M.~K\"orner, and A.~P. Young.
\newblock Universality in three-dimensional ising spin glasses: A monte carlo
  study.
\newblock {\em Phys. Rev. B}, 73:224432, 2006.

\bibitem{machta:08}
J.~Machta, C.~M. Newman, and D.~L. Stein.
\newblock The percolation signature of the spin glass transition.
\newblock {\em Journal of Statistical Physics}, 130(1):113--128, 2008.

\bibitem{hasenbusch:08}
M.~Hasenbusch, A.~Pelissetto, and E.~Vicari.
\newblock {\em J. Stat. Mech.}, 2008:L02001, 2008.

\bibitem{hasenbusch:08b}
Martin Hasenbusch, Andrea Pelissetto, and Ettore Vicari.
\newblock Critical behavior of three-dimensional ising spin glass models.
\newblock {\em Phys. Rev. B}, 78:214205, Dec 2008.

\bibitem{fernandez:09b}
L.~A. Fernandez, V.~Mart{\'i}n-Mayor, S.~Perez-Gaviro, A.~Tarancon, and A.~P.
  Young.
\newblock Phase transition in the three dimensional {H}eisenberg spin glass:
  Finite-size scaling analysis.
\newblock {\em Phys. Rev. B}, 80:024422, 2009.

\bibitem{janus:13}
M.~Baity-Jesi, R.~A. Ba\~{n}os, Andres Cruz, Luis~Antonio Fernandez,
  Jose~Miguel Gil-Narvion, Antonio Gordillo-Guerrero, David Iniguez, Andrea
  Maiorano, F.~Mantovani, Enzo Marinari, Victor Mart\'{i}n-Mayor, Jorge
  Monforte-Garcia, Antonio Mu{\~n}oz~Sudupe, Denis Navarro, Giorgio Parisi,
  Sergio Perez-Gaviro, M.~Pivanti, F.~Ricci-Tersenghi, Juan~Jesus Ruiz-Lorenzo,
  Sebastiano~Fabio Schifano, Beatriz Seoane, Alfonso Tarancon, Raffaele
  Tripiccione, and David Yllanes.
\newblock Critical parameters of the three-dimensional ising spin glass.
\newblock {\em Phys. Rev. B}, 88:224416, {2013}.

\bibitem{seoane:13}
B.~Seoane.
\newblock {\em Spin glasses, the quantum annealing, colloidal glasses and
  crystals: exploring complex free-energy landscapes}.
\newblock PhD thesis, Universidad Complutense de Madrid, January 2013.

\bibitem{yllanes:11}
D.~Yllanes.
\newblock {\em Rugged Free-Energy Landscapes in Disordered Spin Systems}.
\newblock PhD thesis, Universidad Complutense de Madrid, 2011.

\bibitem{pearson:81}
Robert~B Pearson, JL~Richardson, and Doug Toussaint.
\newblock A special purpose machine for monte-carlo simulation.
\newblock Technical Report NSF-ITP-81-139, Inst. Theoretical Physics, Univ.
  California, Santa Barbara, 1981.

\bibitem{condon:85}
JH~Condon and AT~Ogielski.
\newblock Fast special purpose computer for monte carlo simulations in
  statistical physics.
\newblock {\em Review of scientific instruments}, 56(9):1691--1696, 1985.

\bibitem{vandersijs:96}
A.~J. van~der Sijs.
\newblock {RTNN: The New Parallel Machine in Zaragoza}.
\newblock {\em Progress of Theoretical Physics Supplement}, 122:31--40, 01
  1996.

\bibitem{cruz:01}
A.~Cruz, J.~Pech, A.~Tarancon, P.~Tellez, C.~L. Ullod, and C.~Ungil.
\newblock {SUE}: A special purpose computer for spin glass models.
\newblock {\em Comp. Phys. Comm}, 133:165--176, 2001.

\bibitem{janus:06}
F.~Belletti, F.~Mantovani, G.~Poli, S.~F. Schifano, R.~Tripiccione, I.~Campos,
  A.~Cruz, D.~Navarro, S.~Perez-Gaviro, D.~Sciretti, A.~Tarancon, J.~L.
  Velasco, P.~Tellez, L.~A. Fernandez, V.~Mart\'{i}n-Mayor,
  A.~Mu{\~n}oz~Sudupe, S.~Jimenez, A.~Maiorano, E.~Marinari, and J.~J.
  Ruiz-Lorenzo.
\newblock Ianus: And adaptive fpga computer.
\newblock {\em Computing in Science and Engineering}, 8:41, 2006.

\bibitem{janus:08}
F.~Belletti, M.~Cotallo, A.~Cruz, L.~A. Fernandez, A.~Gordillo, A.~Maiorano,
  F.~Mantovani, E.~Marinari, V.~Mart\'{i}n-Mayor, A.~Mu{\~n}oz~Sudupe,
  D.~Navarro, S.~Perez-Gaviro, J.~J. Ruiz-Lorenzo, S.~F. Schifano, D.~Sciretti,
  A.~Tarancon, R.~Tripiccione, and J.~L. Velasco.
\newblock Simulating spin systems on {IANUS}, an {FPGA}-based computer.
\newblock {\em Comp. Phys. Comm.}, 178:208--216, 2008.

\bibitem{janus:12b}
M.~Baity-Jesi, R.~A. Ba\~{n}os, A.~Cruz, L.~A. Fernandez, J.~M. Gil-Narvion,
  A.~Gordillo-Guerrero, M.~Guidetti, D.~Iniguez, A.~Maiorano, F.~Mantovani,
  E.~Marinari, V.~Mart\'{i}n-Mayor, J.~Monforte-Garcia, A.~Munoz~Sudupe,
  D.~Navarro, G.~Parisi, M.~Pivanti, S.~Perez-Gaviro, F.~Ricci-Tersenghi, J.~J.
  Ruiz-Lorenzo, S.~F. Schifano, B.~Seoane, A.~Tarancon, P.~Tellez,
  R.~Tripiccione, and D.~Yllanes.
\newblock {Reconfigurable computing for Monte Carlo simulations: Results and
  prospects of the Janus project}.
\newblock {\em Eur. Phys. J. Special Topics}, {210}:{33}, {AUG} {2012}.

\bibitem{janus:13c}
M.~Baity-Jesi, R.A. Baños, A.~Cruz, L.A. Fernandez, J.M. Gil-Narvion,
  A.~Gordillo-Guerrero, D.~Iñiguez, A.~Maiorano, F.~Mantovani, E.~Marinari,
  V.~Martin-Mayor, J.~Monforte-Garcia, A.~Muñoz Sudupe, D.~Navarro, G.~Parisi,
  S.~Perez-Gaviro, M.~Pivanti, F.~Ricci-Tersenghi, J.J. Ruiz-Lorenzo, S.F.
  Schifano, B.~Seoane, A.~Tarancon, R.~Tripiccione, and D.~Yllanes.
\newblock The janus project: boosting spin-glass simulations using fpgas.
\newblock {\em IFAC Proceedings Volumes}, 46(28):227 -- 232, 2013.
\newblock 12th IFAC Conference on Programmable Devices and Embedded Systems.

\bibitem{janus:14}
M.~Baity-Jesi, R.~A. Ba\~{n}os, Andres Cruz, Luis~Antonio Fernandez,
  Jose~Miguel Gil-Narvion, Antonio Gordillo-Guerrero, David Iniguez, Andrea
  Maiorano, F.~Mantovani, Enzo Marinari, Victor Mart\'{i}n-Mayor, Jorge
  Monforte-Garcia, Antonio Mu{\~n}oz~Sudupe, Denis Navarro, Giorgio Parisi,
  Sergio Perez-Gaviro, M.~Pivanti, F.~Ricci-Tersenghi, Juan~Jesus Ruiz-Lorenzo,
  Sebastiano~Fabio Schifano, Beatriz Seoane, Alfonso Tarancon, Raffaele
  Tripiccione, and David Yllanes.
\newblock Janus {II}: a new generation application-driven computer for
  spin-system simulations.
\newblock {\em Comp. Phys. Comm}, 185:550--559, 2014.

\bibitem{weigel:11}
Martin Weigel.
\newblock Simulating spin models on gpu.
\newblock {\em Computer Physics Communications}, 182(9):1833--1836, 2011.

\bibitem{weigel:12}
Martin Weigel.
\newblock Performance potential for simulating spin models on gpu.
\newblock {\em Journal of Computational Physics}, 231(8):3064--3082, 2012.

\bibitem{fang:14}
Y.~Fang, S.~Feng, K.-M. Tam, Z.~Yun, J.~Moreno, J.~Ramanujam, and M.~Jarrell.
\newblock Parallel tempering simulation of the three-dimensional
  edwards-anderson model with compact asynchronous multispin coding on gpu.
\newblock {\em Comp. Phys. Comm.}, 185:2467–--2478, 2014.

\bibitem{lulli:15}
Matteo Lulli, Giorgio Parisi, and Andrea Pelissetto.
\newblock Out-of-equilibrium finite-size method for critical behavior analyses.
\newblock {\em Phys. Rev. E}, 93:032126, Mar 2016.

\bibitem{newman:99}
M.~E.~J. Newman and G.~T. Barkema.
\newblock {\em {M}onte {C}arlo Methods in Statistical Physics}.
\newblock Clarendon Press, Oxford, 1999.

\bibitem{kawashima:96}
N.~Kawashima and A.~P. Young.
\newblock Phase transition in the three-dimensional
  $\ifmmode\pm\else\textpm\fi{}j$ ising spin glass.
\newblock {\em Phys. Rev. B}, 53:R484--R487, Jan 1996.

\bibitem{isakov:15}
S.V. Isakov, I.N. Zintchenko, T.F. Rønnow, and M.~Troyer.
\newblock Optimised simulated annealing for ising spin glasses.
\newblock {\em Computer Physics Communications}, 192:265--271, 2015.

\bibitem{jorg:08b}
T.~J{\"o}rg, H.~G. Katzgraber, and F.~Krzakala.
\newblock Behavior of {I}sing spin glasses in a magnetic field.
\newblock {\em Phys. Rev. Lett.}, 100:197202, 2008.

\bibitem{fernandez:16b}
L.~A. Fernandez, E.~Marinari, V.~Martin-Mayor, G.~Parisi, and J.~J.
  Ruiz-Lorenzo.
\newblock Universal critical behavior of the two-dimensional ising spin glass.
\newblock {\em Phys. Rev. B}, 94:024402, Jul 2016.

\bibitem{swendsen:86}
Robert~H Swendsen and Jian-Sheng Wang.
\newblock Replica monte carlo simulation of spin-glasses.
\newblock {\em Physical review letters}, 57(21):2607, 1986.

\bibitem{hukushima:96}
K.~Hukushima and K.~Nemoto.
\newblock Exchange monte carlo method and application to spin glass
  simulations.
\newblock {\em J. Phys. Soc. Japan}, 65:1604, 1996.

\bibitem{wang:15b}
Wenlong Wang, Jonathan Machta, and Helmut~G Katzgraber.
\newblock Comparing monte carlo methods for finding ground states of ising spin
  glasses: Population annealing, simulated annealing, and parallel tempering.
\newblock {\em Physical Review E}, 92(1):013303, 2015.

\bibitem{joh:99}
Y.~G. Joh, R.~Orbach, G.~G. Wood, J.~Hammann, and E.~Vincent.
\newblock Extraction of the spin glass correlation length.
\newblock {\em Phys. Rev. Lett.}, 82:438--441, Jan 1999.

\bibitem{nakamae:12}
S.~Nakamae, C.~Crauste-Thibierge, D.~L'H\^ote, E.~Vincent, E.~Dubois,
  V.~Dupuis, and R.~Perzynski.
\newblock Dynamic correlation length growth in superspin glass: Bridging
  experiments and simulations.
\newblock {\em Appl. Phys. Lett.}, 101:242409, 2012.

\bibitem{guchhait:14}
Samaresh Guchhait and Raymond Orbach.
\newblock Direct dynamical evidence for the spin glass lower critical dimension
  $2<d_l<3$.
\newblock {\em Phys. Rev. Lett.}, 112:126401, Mar 2014.

\bibitem{guchhait:17}
Samaresh Guchhait and Raymond~L. Orbach.
\newblock Magnetic field dependence of spin glass free energy barriers.
\newblock {\em Phys. Rev. Lett.}, 118:157203, Apr 2017.

\bibitem{janus:10}
R.~Alvarez~Ba{\~n}os, A.~Cruz, L.~A. Fernandez, J.~M. Gil-Narvion,
  A.~Gordillo-Guerrero, M.~Guidetti, A.~Maiorano, F.~Mantovani, E.~Marinari,
  V.~Mart\'{i}n-Mayor, J.~Monforte-Garcia, A.~Mu{\~n}oz~Sudupe, D.~Navarro,
  G.~Parisi, S.~Perez-Gaviro, J.~J. Ruiz-Lorenzo, S.~F. Schifano, B.~Seoane,
  A.~Tarancon, R.~Tripiccione, and D.~Yllanes.
\newblock Nature of the spin-glass phase at experimental length scales.
\newblock {\em J. Stat. Mech.}, 2010:P06026, 2010.

\bibitem{janus:09b}
F.~Belletti, A.~Cruz, L.~A. Fernandez, A.~Gordillo-Guerrero, M.~Guidetti,
  A.~Maiorano, F.~Mantovani, E.~Marinari, V.~Mart\'{i}n-Mayor, J.~Monforte,
  A.~Mu{\~n}oz~Sudupe, D.~Navarro, G.~Parisi, S.~Perez-Gaviro, J.~J.
  Ruiz-Lorenzo, S.~F. Schifano, D.~Sciretti, A.~Tarancon, R.~Tripiccione, and
  D.~Yllanes.
\newblock An in-depth view of the microscopic dynamics of ising spin glasses at
  fixed temperature.
\newblock {\em J. Stat. Phys.}, 135:1121, 2009.

\bibitem{caracciolo:90}
Sergio Caracciolo, Giorgio Parisi, Stefano Patarnello, and Nicolas Sourlas.
\newblock 3d ising spin-glasses in a magnetic field and mean-field theory.
\newblock {\em EPL (Europhysics Letters)}, 11(8):783, 1990.

\bibitem{altieri2016_RG}
Ada Altieri, Giorgio Parisi, and Tommaso Rizzo.
\newblock Composite operators in cubic field theories and link-overlap
  fluctuations in spin-glass models.
\newblock {\em Physical Review B}, 93(2):024422, 2016.

\bibitem{cooper:82}
F.~Cooper, B.~Freedman, and D.~Preston.
\newblock Solving {$\phi_{1,2}^4$} field theory with {M}onte {C}arlo.
\newblock {\em Nucl. Phys. B}, 210:210, 1982.

\bibitem{janus:08b}
F.~Belletti, M~Cotallo, A.~Cruz, L.~A. Fernandez, A.~Gordillo-Guerrero,
  M.~Guidetti, A.~Maiorano, F.~Mantovani, E.~Marinari, V.~Mart\'{i}n-Mayor,
  A.~M. Sudupe, D.~Navarro, G.~Parisi, S.~Perez-Gaviro, J.~J. Ruiz-Lorenzo,
  S.~F. Schifano, D.~Sciretti, A.~Tarancon, R.~Tripiccione, J.~L. Velasco, and
  D.~Yllanes.
\newblock Nonequilibrium spin-glass dynamics from picoseconds to one tenth of a
  second.
\newblock {\em Phys. Rev. Lett.}, 101:157201, 2008.

\bibitem{baityjesi:16}
Marco Baity-Jesi.
\newblock {\em Spin Glasses: Criticality and Energy Landscapes}.
\newblock Springer Theses. Springer International Publishing, 1 edition, 2016.

\bibitem{janus:23}
M~Baity-Jesi, E~Calore, A~Cruz, LA~Fernandez, JM~Gil-Narvion, I~Pemartin,
  A~Gordillo-Guerrero, D~I{\~n}iguez, A~Maiorano, E~Marinari, et~al.
\newblock Memory and rejuvenation in spin glasses: aging systems are ruled by
  more than one length scale.
\newblock {\em arXiv preprint arXiv:2207.06207}, 2022.

\bibitem{janus:17b}
M.~Baity-Jesi, E.~Calore, A.~Cruz, L.~A. Fernandez, J.~M. Gil-Narvion,
  A.~Gordillo-Guerrero, D.~I\~niguez, A.~Maiorano, E.~Marinari,
  V.~Martin-Mayor, J.~Monforte-Garcia, A.~Mu\~noz Sudupe, D.~Navarro,
  G.~Parisi, S.~Perez-Gaviro, F.~Ricci-Tersenghi, J.~J. Ruiz-Lorenzo, S.~F.
  Schifano, B.~Seoane, A.~Tarancon, R.~Tripiccione, and D.~Yllanes.
\newblock Matching microscopic and macroscopic responses in glasses.
\newblock {\em Phys. Rev. Lett.}, 118:157202, Apr 2017.

\bibitem{fisher:72}
ME~Fisher.
\newblock Critical phenomena, proc. 51st enrico fermi summer school, varena,
  1972.

\bibitem{nightingale:76}
M.P Nightingale.
\newblock Scaling theory and finite systems.
\newblock {\em Physica A: Statistical Mechanics and its Applications},
  83(3):561 -- 572, 1976.

\bibitem{binder:81}
K.~Binder.
\newblock {Finite size scaling analysis of ising model block distribution
  functions}.
\newblock {\em Z. Phys. B -- Condensed Matter}, 43:119--140, 1981.

\bibitem{janus:12}
R.~A. Ba\~{n}os, Andres Cruz, Luis~Antonio Fernandez, Jose~Miguel Gil-Narvion,
  Antonio Gordillo-Guerrero, Marco Guidetti, David Iniguez, Andrea Maiorano,
  Enzo Marinari, Victor Mart\'{i}n-Mayor, Jorge Monforte-Garcia, Antonio
  Mu{\~n}oz~Sudupe, Denis Navarro, Giorgio Parisi, Sergio Perez-Gaviro,
  Juan~Jesus Ruiz-Lorenzo, Sebastiano~Fabio Schifano, Beatriz Seoane, Alfonso
  Tarancon, Pedro Tellez, Raffaele Tripiccione, and David Yllanes.
\newblock {Thermodynamic glass transition in a spin glass without time-reversal
  symmetry}.
\newblock {\em Proc. Natl. Acad. Sci. USA}, {109}:6452, {2012}.

\bibitem{janus:14c}
M.~Baity-Jesi, R.~A. Ba\~{n}os, Andres Cruz, Luis~Antonio Fernandez,
  Jose~Miguel Gil-Narvion, Antonio Gordillo-Guerrero, David Iniguez, Andrea
  Maiorano, Mantovani F., Enzo Marinari, Victor Mart\'{i}n-Mayor, Jorge
  Monforte-Garcia, Antonio Mu{\~n}oz~Sudupe, Denis Navarro, Giorgio Parisi,
  Sergio Perez-Gaviro, M.~Pivanti, F.~Ricci-Tersenghi, Juan~Jesus Ruiz-Lorenzo,
  Sebastiano~Fabio Schifano, Beatriz Seoane, Alfonso Tarancon, Raffaele
  Tripiccione, and David Yllanes.
\newblock The three dimensional {I}sing spin glass in an external magnetic
  field: the role of the silent majority.
\newblock {\em J. Stat. Mech.}, 2014:P05014, {2014}.

\bibitem{fisher:87}
D.~S. Fisher and D.~A. Huse.
\newblock Absence of many states in realistic spin glasses.
\newblock {\em J. Phys. A: Math. Gen.}, 20:{L}1005, 1987.

\bibitem{huse:87}
David~A. Huse and Daniel~S. Fisher.
\newblock Dynamics of droplet fluctuations in pure and random ising systems.
\newblock {\em Phys. Rev. B}, 35:6841--6846, May 1987.

\bibitem{fisher:88}
D.~S. Fisher and D.~A. Huse.
\newblock Nonequilibrium dynamics of spin glasses.
\newblock {\em Phys. Rev. B}, 38:373, 1988.

\bibitem{fisher:88b}
D.~S. Fisher and D.~A. Huse.
\newblock Equilibrium behavior of the spin-glass ordered phase.
\newblock {\em Phys. Rev. B}, 38:386, 1988.

\bibitem{migdal:75}
A.A. Migdal.
\newblock Phase transitions in gauge and spin-lattice systems.
\newblock {\em Zhurnal Eksperimentalnoi i teoreticheskoi fiziki}, 1975.

\bibitem{kadanoff:76}
L.P. Kadanoff.
\newblock Notes on migdal's recursion formulas.
\newblock {\em Annals of Physics}, 100:359--394, 1976.

\bibitem{huang:87}
K.~Huang.
\newblock {\em Statistical Mechanics}.
\newblock John Wiley and Sons, Hoboken, NJ, second edition, 1987.

\bibitem{parisi:96}
G.~Parisi.
\newblock Recent rigorous results support the predictions of spontaneously
  broken replica symmetry for realistic spin glasses.
\newblock Reply to \cite{newman:96c}., 1996.

\bibitem{marinari:00}
E.~Marinari, G.~Parisi, F.~Ricci-Tersenghi, J.~J. Ruiz-Lorenzo, and F.~Zuliani.
\newblock Replica symmetry breaking in short-range spin glasses: Theoretical
  foundations and numerical evidences.
\newblock {\em J. Stat. Phys.}, 98:973, 2000.

\bibitem{parisi:12}
G.~Parisi and T.~Temesv\'ari.
\newblock Replica symmetry breaking in and around six dimensions.
\newblock {\em Nucl. Phys. B}, 858:293, 2012.

\bibitem{moore:11}
M.~A. Moore and A.~J. Bray.
\newblock Disappearance of the de {A}lmeida-{T}houless line in six dimensions.
\newblock {\em Phys. Rev. B}, 83:224408, 2011.

\bibitem{yeo:12}
J.~Yeo and M.~A. Moore.
\newblock Origin of the growing length scale in mp-spin glass models.
\newblock {\em Phys. Rev. E}, 86:052501, 2012.

\bibitem{yucesoy:12}
B.~Yucesoy, Helmut~G. Katzgraber, and J.~Machta.
\newblock Evidence of non-mean-field-like low-temperature behavior in the
  edwards-anderson spin-glass model.
\newblock {\em Phys. Rev. Lett.}, 109:177204, Oct 2012.

\bibitem{yucesoy:13}
B.~Yucesoy, Helmut~G. Katzgraber, and J.~Machta.
\newblock Reply to comment.
\newblock {\em Phys. Rev. Lett.}, 110:219702, 2013.

\bibitem{billoire:13}
A.~Billoire, L.~A. Fernandez, A.~Maiorano, E.~Marinari, V.~Mart{\'i}n-Mayor,
  G.~Parisi, F.~Ricci-Tersenghi, J.~J. Ruiz-Lorenzo, and D.~Yllanes.
\newblock Comment on ``evidence of non-mean-field-like low-temperature behavior
  in the edwards-anderson spin-glass model''.
\newblock {\em Phys. Rev. Lett.}, 110:219701, 2013.

\bibitem{ruizlorenzo:20}
Juan~J Ruiz-Lorenzo.
\newblock Nature of the spin glass phase in finite dimensional (ising) spin
  glasses.
\newblock In {\em Order, Disorder and Criticality: Advanced Problems of Phase
  Transition Theory}, pages 1--52. World Scientific, 2020.

\bibitem{holler:20}
J.~H\"oller and N.~Read.
\newblock One-step replica-symmetry-breaking phase below the de
  almeida--thouless line in low-dimensional spin glasses.
\newblock {\em Phys. Rev. E}, 101:042114, Apr 2020.

\bibitem{moore:21}
M.~A. Moore.
\newblock Droplet-scaling versus replica symmetry breaking debate in spin
  glasses revisited.
\newblock {\em Phys. Rev. E}, 103:062111, Jun 2021.

\bibitem{newman:22}
C.~M. Newman and D.~L. Stein.
\newblock Ground-state stability and the nature of the spin glass phase.
\newblock {\em Phys. Rev. E}, 105:044132, Apr 2022.

\bibitem{martin-mayor:22}
V~Martin-Mayor, JJ~Ruiz-Lorenzo, B~Seoane, and AP~Young.
\newblock Numerical simulations and replica symmetry breaking.
\newblock In World Scientific, editor, {\em Spin Glass Theory \& Far Beyond -
  Replica Symmetry Breaking after 40 Years}. 2022.

\bibitem{janus:11}
R.~A. Ba\~nos, A.~Cruz, L.~A. Fernandez, J.~M. Gil-Narvion,
  A.~Gordillo-Guerrero, M.~Guidetti, D.~I\~niguez, A.~Maiorano, F.~Mantovani,
  E.~Marinari, V.~Mart\'{i}n-Mayor, J.~Monforte-Garcia, A.~Mu\~noz Sudupe,
  D.~Navarro, G.~Parisi, S.~Perez-Gaviro, F.~Ricci-Tersenghi, J.~J.
  Ruiz-Lorenzo, S.~F. Schifano, B.~Seoane, A.~Taranc\'on, R.~Tripiccione, and
  D.~Yllanes.
\newblock Sample-to-sample fluctuations of the overlap distributions in the
  three-dimensional edwards-anderson spin glass.
\newblock {\em Phys. Rev. B}, 84:174209, Nov 2011.

\bibitem{janus:14b}
M.~Baity-Jesi, R.~A. Ba\~{n}os, Andres Cruz, Luis~Antonio Fernandez,
  Jose~Miguel Gil-Narvion, Antonio Gordillo-Guerrero, David Iniguez, Andrea
  Maiorano, Mantovani F., Enzo Marinari, Victor Mart\'{i}n-Mayor, Jorge
  Monforte-Garcia, Antonio Mu{\~n}oz~Sudupe, Denis Navarro, Giorgio Parisi,
  Sergio Perez-Gaviro, M.~Pivanti, F.~Ricci-Tersenghi, Juan~Jesus Ruiz-Lorenzo,
  Sebastiano~Fabio Schifano, Beatriz Seoane, Alfonso Tarancon, Raffaele
  Tripiccione, and David Yllanes.
\newblock {Dynamical Transition in the D=3 Edwards-Anderson spin glass in an
  external magnetic field}.
\newblock {\em Phys. Rev. E}, 89:032140, {2014}.

\bibitem{vedula2023}
Bharadwaj Vedula, MA~Moore, and Auditya Sharma.
\newblock Study of the de almeida-thouless (at) line in the one-dimensional
  diluted power-law xy spin glass.
\newblock {\em arXiv preprint arXiv:2301.03615}, 2023.

\bibitem{angelini:22}
Maria~Chiara Angelini, Carlo Lucibello, Giorgio Parisi, Gianmarco Perrupato,
  Federico Ricci-Tersenghi, and Tommaso Rizzo.
\newblock Unexpected upper critical dimension for spin glass models in a field
  predicted by the loop expansion around the bethe solution at zero
  temperature.
\newblock {\em Phys. Rev. Lett.}, 128:075702, Feb 2022.

\bibitem{temesvari:23}
Tam{\'a}s Temesv{\'a}ri and Imre Kondor.
\newblock Field theory for the almeida-thouless transition.
\newblock 2023.

\bibitem{temesvari:08}
T.~Temesv{\'a}ri.
\newblock {\em Phys. Rev. B}, 78:220401, 2008.

\bibitem{charbonneau:17b}
Patrick Charbonneau and Sho Yaida.
\newblock Nontrivial critical fixed point for replica-symmetry-breaking
  transitions.
\newblock {\em Physical review letters}, 118(21):215701, 2017.

\bibitem{charbonneau:19}
Patrick Charbonneau, Yi~Hu, Archishman Raju, James~P Sethna, and Sho Yaida.
\newblock Morphology of renormalization-group flow for the de
  almeida--thouless--gardner universality class.
\newblock {\em Physical Review E}, 99(2):022132, 2019.

\bibitem{janus:18}
M.~Baity-Jesi, E.~Calore, A.~Cruz, L.~A. Fernandez, J.~M. Gil-Narvion,
  A.~Gordillo-Guerrero, D.~I\~niguez, A.~Maiorano, E.~Marinari,
  V.~Martin-Mayor, J.~Moreno-Gordo, A.~Mu\~noz Sudupe, D.~Navarro, G.~Parisi,
  S.~Perez-Gaviro, F.~Ricci-Tersenghi, J.~J. Ruiz-Lorenzo, S.~F. Schifano,
  B.~Seoane, A.~Tarancon, R.~Tripiccione, and D.~Yllanes.
\newblock Aging rate of spin glasses from simulations matches experiments.
\newblock {\em Phys. Rev. Lett.}, 120:267203, Jun 2018.

\bibitem{zhai:19}
Qiang Zhai, V~Martin-Mayor, Deborah~L Schlagel, Gregory~G Kenning, and
  Raymond~L Orbach.
\newblock Slowing down of spin glass correlation length growth: Simulations
  meet experiments.
\newblock {\em Physical Review B}, 100(9):094202, 2019.

\bibitem{paga:22}
Ilaria Paga.
\newblock From glassy bulk systems to spin-glass films: simulations meet
  experiments.
\newblock {\em Ene}, 11:44, 2022.

\bibitem{palassini:00}
M.~Palassini and A.~P. Young.
\newblock {\em Phys. Rev. Lett.}, 85:3017, 2000.

\bibitem{krzakala:00}
F.~Krzakala and O.~C. Martin.
\newblock {\em Phys. Rev. Lett.}, 85:3013, 2000.

\bibitem{palassini:03}
M.~Palassini, F.~Liers, M.~Juenger, and A.~P. Young.
\newblock {\em Phys. Rev. B}, 68:064413, 2003.

\bibitem{chatterjee:23}
Sourav Chatterjee.
\newblock Spin glass phase at zero temperature in the edwards-anderson model.
\newblock {\em arXiv preprint arXiv:2301.04112}, 2023.

\bibitem{cizeau:93}
Pierre Cizeau and Jean-Philippe Bouchaud.
\newblock Mean field theory of dilute spin-glasses with power-law interactions.
\newblock {\em Journal of Physics A: Mathematical and General}, 26(5):L187,
  1993.

\bibitem{janzen:10}
K.~Janzen, A.~Engel, and M.~M\'ezard.
\newblock Thermodynamics of the l\'evy spin glass.
\newblock {\em Phys. Rev. E}, 82:021127, Aug 2010.

\bibitem{andresen:11}
Juan~Carlos Andresen, Katharina Janzen, and Helmut~G Katzgraber.
\newblock Critical behavior and universality in l{\'e}vy spin glasses.
\newblock {\em Physical Review B}, 83(17):174427, 2011.

\bibitem{katzgraber:03}
H.~Katzgraber and A.~P. Young.
\newblock {\em Phys. Rev. B}, 67:134410, 2003.

\bibitem{katzgraber:05}
Helmut~G. Katzgraber and A.~P. Young.
\newblock Probing the almeida-thouless line away from the mean-field model.
\newblock {\em Phys. Rev. B}, 72:184416, Nov 2005.

\bibitem{banos:12}
R.~A. Ba\~nos, L.~A. Fernandez, V.~Martin-Mayor, and A.~P. Young.
\newblock Correspondence between long-range and short-range spin glasses.
\newblock {\em Phys. Rev. B}, 86:134416, Oct 2012.

\bibitem{leuzzi:13}
L.~Leuzzi and G.~Parisi.
\newblock Long-range random-field ising model: Phase transition threshold and
  equivalence of short and long ranges.
\newblock {\em Phys. Rev. B}, 88:224204, 2013.

\bibitem{wittmann:16}
Matthew Wittmann and A~P Young.
\newblock The connection between statics and dynamics of spin glasses.
\newblock {\em J. Stat. Mech.: Theory Exp}, 2016(1):013301, 2016.

\bibitem{cugliandolo:94}
L.~F. Cugliandolo, J.~Kurchan, and G.~Parisi.
\newblock {\em J. Phys. (France)}, 4:1641, 1994.

\bibitem{barrat:98}
A~Barrat and S~Franz.
\newblock Basins of attraction of metastable states of the spherical p-spin
  model.
\newblock {\em J. Phys. A}, 31(6):L119, 1998.

\bibitem{dealmeida:78b}
J.R.L. de~Almeida, R.C. Jones, J.M. Kosterlitz, and D.J. Thouless.
\newblock The infinite-ranged spin glass with m-component spins.
\newblock {\em Journal of Physics C: Solid State Physics}, 11(21):L871, 1978.

\bibitem{hastings:00}
M.~B. Hastings.
\newblock {\em J. Stat. Phys.}, 99:171, 2000.

\bibitem{baityjesi:15}
M.~Baity-Jesi and G.~Parisi.
\newblock Inherent structures in m-component spin glasses.
\newblock {\em Phys. Rev. B}, 91(13):134203, April 2015.

\bibitem{aspelmeier:04b}
T~Aspelmeier and MA~Moore.
\newblock Generalized bose-einstein phase transition in large-m component spin
  glasses.
\newblock {\em Physical review letters}, 92(7):077201, 2004.

\bibitem{moore:12}
M.~A. Moore.
\newblock {\em Phys. Rev. E}, 86:052501, 2012.

\bibitem{lee:05}
L.~W. Lee, A.~Dhar, and A.~P. Young.
\newblock {\em Phys. Rev. E}, 71:036146, 2005.

\bibitem{barrat:97}
Alain Barrat, Silvio Franz, and Giorgio Parisi.
\newblock Temperature evolution and bifurcations of metastable states in
  mean-field spin glasses, with connections with structural glasses.
\newblock {\em Journal of Physics A: Mathematical and General}, 30(16):5593,
  1997.

\bibitem{crisanti:04}
A.~Crisanti and L.~Leuzzi.
\newblock Spherical $2+p$ spin-glass model: An exactly solvable model for glass
  to spin-glass transition.
\newblock {\em Phys. Rev. Lett.}, 93:217203, Nov 2004.

\bibitem{folena:20}
Giampaolo Folena, Silvio Franz, and Federico Ricci-Tersenghi.
\newblock Rethinking mean-field glassy dynamics and its relation with the
  energy landscape: The surprising case of the spherical mixed $p$-spin model.
\newblock {\em Phys. Rev. X}, 10:031045, Aug 2020.

\bibitem{folena:21}
Giampaolo Folena, Silvio Franz, and Federico Ricci-Tersenghi.
\newblock Gradient descent dynamics in the mixed p-spin spherical model:
  finite-size simulations and comparison with mean-field integration.
\newblock {\em Journal of Statistical Mechanics: Theory and Experiment},
  2021(3):033302, 2021.

\bibitem{folena:20b}
Giampaolo Folena.
\newblock {\em The mixed p-spin model: selecting, following and losing states}.
\newblock PhD thesis, Universit{\'e} Paris-Saclay; Universit{\`a} degli studi
  La Sapienza (Rome), 2020.

\bibitem{debenedetti:21}
Pablo~G Debenedetti.
\newblock Metastable liquids.
\newblock In {\em Metastable Liquids}. Princeton university press, 2021.

\bibitem{gardner:89}
E~Gardner and B~Derrida.
\newblock Three unfinished works on the optimal storage capacity of networks.
\newblock {\em Journal of Physics A: Mathematical and General}, 22(12):1983,
  jun 1989.

\bibitem{mezard:02}
Marc M{\'e}zard, Giorgio Parisi, and Riccardo Zecchina.
\newblock Analytic and algorithmic solution of random satisfiability problems.
\newblock {\em Science}, 297(5582):812--815, 2002.

\bibitem{baldassi:16}
Carlo Baldassi, Christian Borgs, Jennifer~T. Chayes, Alessandro Ingrosso, Carlo
  Lucibello, Luca Saglietti, and Riccardo Zecchina.
\newblock Unreasonable effectiveness of learning neural networks: {From}
  accessible states and robust ensembles to basic algorithmic schemes.
\newblock {\em Proceedings of the National Academy of Sciences},
  113(48):E7655--E7662, November 2016.

\bibitem{mannelli2020}
Stefano~Sarao Mannelli, Giulio Biroli, Chiara Cammarota, Florent Krzakala,
  Pierfrancesco Urbani, and Lenka Zdeborov{\'a}.
\newblock Marvels and pitfalls of the langevin algorithm in noisy
  high-dimensional inference.
\newblock {\em Physical Review X}, 10(1):011057, 2020.

\bibitem{chaudhari:15}
Pratik Chaudhari and Stefano Soatto.
\newblock On the energy landscape of deep networks.
\newblock {\em arXiv:1511.06485}, 2015.

\bibitem{franz2015}
Silvio Franz, Giorgio Parisi, Pierfrancesco Urbani, and Francesco Zamponi.
\newblock Universal spectrum of normal modes in low-temperature glasses.
\newblock {\em Proceedings of the National Academy of Sciences},
  112(47):14539--14544, 2015.

\bibitem{antenucci2019}
Fabrizio Antenucci, Silvio Franz, Pierfrancesco Urbani, and Lenka
  Zdeborov{\'a}.
\newblock Glassy nature of the hard phase in inference problems.
\newblock {\em Physical Review X}, 9(1):011020, 2019.

\bibitem{ros:19}
Valentina Ros, Gerard Ben~Arous, Giulio Biroli, and Chiara Cammarota.
\newblock Complex energy landscapes in spiked-tensor and simple glassy models:
  Ruggedness, arrangements of local minima, and phase transitions.
\newblock {\em Physical Review X}, 9(1):011003, 2019.

\bibitem{ros:19b}
Valentina Ros, Giulio Biroli, and Chiara Cammarota.
\newblock Complexity of energy barriers in mean-field glassy systems.
\newblock {\em EPL (Europhysics Letters)}, 126(2):20003, 2019.

\bibitem{angelani:06}
Luca Angelani, Claudio Conti, Giancarlo Ruocco, and Francesco Zamponi.
\newblock Glassy behavior of light in random lasers.
\newblock {\em Physical Review B}, 74(10):104207, 2006.

\bibitem{antenucci:15}
Fabrizio Antenucci, Andrea Crisanti, and Luca Leuzzi.
\newblock The glassy random laser: replica symmetry breaking in the intensity
  fluctuations of emission spectra.
\newblock {\em Scientific reports}, 5(1):1--11, 2015.

\bibitem{galluccio:98}
Stefano Galluccio, Jean-Philippe Bouchaud, and Marc Potters.
\newblock Rational decisions, random matrices and spin glasses.
\newblock {\em Physica A: Statistical Mechanics and its Applications},
  259(3-4):449--456, 1998.

\bibitem{garnier:2021}
J{\'e}r{\^o}me Garnier-Brun, Michael Benzaquen, Stefano Ciliberti, and
  Jean-Philippe Bouchaud.
\newblock A new spin on optimal portfolios and ecological equilibria.
\newblock {\em Journal of Statistical Mechanics: Theory and Experiment},
  2021(9):093408, 2021.

\bibitem{dessertaine:22}
Th{\'e}o Dessertaine, Jos{\'e} Moran, Michael Benzaquen, and Jean-Philippe
  Bouchaud.
\newblock Out-of-equilibrium dynamics and excess volatility in firm networks.
\newblock {\em Journal of Economic Dynamics and Control}, 138:104362, 2022.

\bibitem{charbonneau:14}
P.~Charbonneau, J.~Kurchan, G.~Parisi, P.~Urbani, and F.~Zamponi.
\newblock Fractal free energy landscapes in structural glasses.
\newblock {\em Nature Communications}, 5:3725, 2014.

\bibitem{rainone:15}
Corrado Rainone, Pierfrancesco Urbani, Hajime Yoshino, and Francesco Zamponi.
\newblock Following the evolution of hard sphere glasses in infinite dimensions
  under external perturbations: Compression and shear strain.
\newblock {\em Physical review letters}, 114(1):015701, 2015.

\bibitem{altieri2018microscopic}
Ada Altieri, Pierfrancesco Urbani, and Francesco Zamponi.
\newblock Microscopic theory of two-step yielding in attractive colloids.
\newblock {\em Physical review letters}, 121(18):185503, 2018.

\bibitem{altieri2019jamming}
Ada Altieri.
\newblock The jamming transition.
\newblock In {\em Jamming and Glass Transitions}, pages 45--64. Springer, 2019.

\bibitem{parisi2020}
Giorgio Parisi, Pierfrancesco Urbani, and Francesco Zamponi.
\newblock {\em Theory of simple glasses: exact solutions in infinite
  dimensions}.
\newblock Cambridge University Press, 2020.

\bibitem{bunin2017ecological}
Guy Bunin.
\newblock Ecological communities with lotka-volterra dynamics.
\newblock {\em Physical Review E}, 95(4):042414, 2017.

\bibitem{biroli:18}
Giulio Biroli, Guy Bunin, and Chiara Cammarota.
\newblock Marginally stable equilibria in critical ecosystems.
\newblock {\em New Journal of Physics}, 20(8):083051, 2018.

\bibitem{altieri2019}
Ada Altieri and Silvio Franz.
\newblock Constraint satisfaction mechanisms for marginal stability and
  criticality in large ecosystems.
\newblock {\em Physical Review E}, 99(1):010401, 2019.

\bibitem{altieri:21}
Ada Altieri, Felix Roy, Chiara Cammarota, and Giulio Biroli.
\newblock Properties of equilibria and glassy phases of the random
  lotka-volterra model with demographic noise.
\newblock {\em Physical Review Letters}, 126(25):258301, 2021.

\bibitem{Altieri2022}
Ada Altieri and Giulio Biroli.
\newblock Effects of intraspecific cooperative interactions in large
  ecosystems.
\newblock {\em SciPost Physics}, 12(1):013, 2022.

\bibitem{roy:20}
Felix Roy, Matthieu Barbier, Giulio Biroli, and Guy Bunin.
\newblock Complex interactions can create persistent fluctuations in
  high-diversity ecosystems.
\newblock {\em PLoS computational biology}, 16(5):e1007827, 2020.

\bibitem{lorenzana:22}
Giulia~Garcia Lorenzana and Ada Altieri.
\newblock Well-mixed lotka-volterra model with random strongly competitive
  interactions.
\newblock {\em Physical Review E}, 105(2):024307, 2022.

\bibitem{altieri2022glassy}
Ada Altieri.
\newblock Glassy features and complex dynamics in ecological systems.
\newblock {\em arXiv preprint arXiv:2208.14956}, 2022.

\bibitem{ros:22}
Valentina Ros, Felix Roy, Giulio Biroli, Guy Bunin, and Ari~M Turner.
\newblock Generalized lotka-volterra equations with random, non-reciprocal
  interactions: the typical number of equilibria.
\newblock 2022.

\bibitem{janus:19}
Marco Baity-Jesi, Enrico Calore, Andres Cruz, Luis~Antonio Fernandez,
  Jos{\'e}~Miguel Gil-Narvi{\'o}n, Antonio Gordillo-Guerrero, David
  I{\~n}iguez, Antonio Lasanta, Andrea Maiorano, Enzo Marinari, Victor
  Martin-Mayor, Javier Moreno-Gordo, Antonio Mu{\~n}oz~Sudupe, Denis Navarro,
  Giorgio Parisi, Sergio Perez-Gaviro, Federico Ricci-Tersenghi, Juan~Jesus
  Ruiz-Lorenzo, Sebastiano~Fabio Schifano, Beatriz Seoane, Alfonso
  Taranc{\'o}n, Raffaele Tripiccione, and David Yllanes.
\newblock The mpemba effect in spin glasses is a persistent memory effect.
\newblock {\em Proceedings of the National Academy of Sciences}, 2019.

\bibitem{benarous:02}
G\'erard Ben~Arous, Anton Bovier, and V\'eronique Gayrard.
\newblock Aging in the random energy model.
\newblock {\em Phys. Rev. Lett.}, 88:087201, Feb 2002.

\bibitem{cammarota:15}
Chiara Cammarota and Enzo Marinari.
\newblock Spontaneous energy-barrier formation in entropy-driven glassy
  dynamics.
\newblock {\em Phys. Rev. E}, 92:010301(R), 2015.

\bibitem{gayrard:16}
V{\'e}ronique Gayrard.
\newblock Convergence of clock processes and aging in metropolis dynamics of a
  truncated {REM}.
\newblock {\em Annales Henri Poincar{\'e}}, 17(3):537--614, 2016.

\bibitem{baityjesi:18}
M.~Baity-Jesi, G.~Biroli, and C.~Cammarota.
\newblock Activated aging dynamics and effective trap model description in the
  random energy model.
\newblock {\em J. Stat. Mech.: Theory Exp}, (1):013301, 2018.

\bibitem{hartarsky:19}
Ivailo Hartarsky, Marco Baity-Jesi, Riccardo Ravasio, Alain Billoire, and
  Giulio Biroli.
\newblock Maximum-energy records in glassy energy landscapes.
\newblock {\em J. Stat. Mech.: Theory Exp}, 2019(9):093302, sep 2019.

\bibitem{gayrard:19}
V{\'e}ronique Gayrard and Lisa Hartung.
\newblock Dynamic phase diagram of the rem.
\newblock In V{\'e}ronique Gayrard, Louis-Pierre Arguin, Nicola Kistler, and
  Irina Kourkova, editors, {\em Statistical Mechanics of Classical and
  Disordered Systems}, pages 111--170, Cham, 2019. Springer International
  Publishing.

\bibitem{stariolo:19}
Daniel~A. Stariolo and Leticia~F. Cugliandolo.
\newblock Activated dynamics of the ising p-spin disordered model with finite
  number of variables.
\newblock {\em {EPL} (Europhysics Letters)}, 127(1):16002, aug 2019.

\bibitem{stariolo:20}
Daniel~A. Stariolo and Leticia~F. Cugliandolo.
\newblock Barriers, trapping times, and overlaps between local minima in the
  dynamics of the disordered ising $p$-spin model.
\newblock {\em Phys. Rev. E}, 102:022126, Aug 2020.

\bibitem{ros:20}
Valentina Ros.
\newblock Distribution of rare saddles in the p-spin energy landscape.
\newblock {\em Journal of Physics A: Mathematical and Theoretical},
  53(12):125002, 2020.

\bibitem{carbone:20}
Matthew~R. Carbone, Valerio Astuti, and Marco Baity-Jesi.
\newblock Effective traplike activated dynamics in a continuous landscape.
\newblock {\em Phys. Rev. E}, 101:052304, May 2020.

\bibitem{ros:21}
Valentina Ros, Giulio Biroli, and Chiara Cammarota.
\newblock Dynamical instantons and activated processes in mean-field glass
  models.
\newblock {\em SciPost Physics}, 10(1):002, 2021.

\bibitem{rizzo:21}
Tommaso Rizzo.
\newblock Path integral approach unveils role of complex energy landscape for
  activated dynamics of glassy systems.
\newblock {\em Physical Review B}, 104(9):094203, 2021.

\bibitem{carbone:22}
Matthew~R Carbone and Marco Baity-Jesi.
\newblock Competition between energy-and entropy-driven activation in glasses.
\newblock {\em Physical Review E}, 106(2):024603, 2022.

\bibitem{fernandez:13}
L.~A. Fernandez, V.~Mart\'{i}n-Mayor, G.~Parisi, and B.~Seoane.
\newblock Temperature chaos in 3d ising spin glasses is driven by rare events.
\newblock {\em EPL}, 103(6):67003, 2013.

\bibitem{billoire:14}
Alain Billoire.
\newblock Rare events analysis of temperature chaos in the
  sherrington–kirkpatrick model.
\newblock {\em J. Stat. Mech.}, 2014(4):P04016, 2014.

\bibitem{guchhait:15b}
Samaresh Guchhait and Raymond~L. Orbach.
\newblock Temperature chaos in a ge:mn thin-film spin glass.
\newblock {\em Phys. Rev. B}, 92:214418, Dec 2015.

\bibitem{fernandez:16}
Luis~Antonio Fernandez, Enzo Marinari, V{\'\i}ctor Mart{\'\i}n-Mayor, Giorgio
  Parisi, and David Yllanes.
\newblock Temperature chaos is a non-local effect.
\newblock {\em Journal of Statistical Mechanics: Theory and Experiment},
  2016(12):123301, 2016.

\bibitem{janus:21b}
Marco Baity-Jesi, Enrico Calore, Andr{\'e}s Cruz, Luis~Antonio Fernandez,
  Jos{\'e}~Miguel Gil-Narvion, Isidoro Gonzalez-Adalid~Pemartin, Antonio
  Gordillo-Guerrero, David I{\~n}iguez, Andrea Maiorano, Enzo Marinari, Victor
  Martin-Mayor, Javier Moreno-Gordo, Antonio Mu{\~n}oz~Sudupe, Denis Navarro,
  Giorgio Parisi, Sergio Perez-Gaviro, Federico Ricci-Tersenghi, Juan~Jesus
  Ruiz-Lorenzo, Sebastiano~Fabio Schifano, Beatriz Seoane, Alfonso
  Taranc{\'o}n, Raffaele Tripiccione, and David Yllanes.
\newblock Temperature chaos is present in off-equilibrium spin-glass dynamics.
\newblock {\em Communications Physics}, 4(1):1--7, 2021.

\bibitem{aizenman:90}
Michael Aizenman and Jan Wehr.
\newblock Rounding effects of quenched randomness on first-order phase
  transitions.
\newblock {\em Communications in Mathematical Physics}, 130(3):489--528, 1990.

\bibitem{newman:92}
C.~M. Newman and D.~L. Stein.
\newblock Multiple states and thermodynamic limits in short-ranged ising
  spin-glass models.
\newblock {\em Phys. Rev. B}, 46:973--982, Jul 1992.

\bibitem{billoire:17}
Alain Billoire, LA~Fernandez, Andrea Maiorano, Enzo Marinari, V{\'\i}ctor
  Martin-Mayor, Javier Moreno-Gordo, Giorgio Parisi, Federico Ricci-Tersenghi,
  and Juan~Jes{\'u}s Ruiz-Lorenzo.
\newblock Numerical construction of the aizenman-wehr metastate.
\newblock {\em Physical Review Letters}, 119(3):037203, 2017.

\bibitem{newman:23}
CM~Newman, N~Read, and DL~Stein.
\newblock Metastates and replica symmetry breaking.
\newblock 2022.

\bibitem{kirkpatrick:78}
Scott Kirkpatrick and David Sherrington.
\newblock Infinite-ranged models of spin-glasses.
\newblock {\em Phys. Rev. B}, 17:4384--4403, Jun 1978.

\bibitem{tarzia:07}
M~Tarzia and MA~Moore.
\newblock Glass phenomenology from the connection to spin glasses.
\newblock {\em Physical Review E}, 75(3):031502, 2007.

\bibitem{baityjesi:15b}
M.~Baity-Jesi, V.~Mart\'{\i}n-Mayor, G.~Parisi, and S.~Perez-Gaviro.
\newblock Soft modes, localization, and two-level systems in spin glasses.
\newblock {\em Phys. Rev. Lett.}, 115:267205, Dec 2015.

\bibitem{choromanska:15}
A.~Choromanska, M.~Henaff, G.~Ben~Arous, and Y.~LeCun.
\newblock The loss surfaces of multilayer networks.
\newblock {\em Proceedings of Machine Learning Research}, 38:192--204, 2015.

\bibitem{kawaguchi:16}
Kenji Kawaguchi.
\newblock Deep learning without poor local minima.
\newblock In D.~D. Lee, M.~Sugiyama, U.~V. Luxburg, I.~Guyon, and R.~Garnett,
  editors, {\em Advances in Neural Information Processing Systems 29}, pages
  586--594. Curran Associates, Inc., 2016.

\bibitem{rizzo:16}
Tommaso Rizzo.
\newblock The glass crossover from mean-field spin-glasses to supercooled
  liquids.
\newblock {\em Philosophical Magazine}, 96(7-9):636--647, 2016.

\bibitem{baityjesi:18b2}
Marco Baity-Jesi, Levent Sagun, Mario Geiger, Stefano Spigler, G{\'{e}}rard~Ben
  Arous, Chiara Cammarota, Yann LeCun, Matthieu Wyart, and Giulio Biroli.
\newblock Comparing dynamics: deep neural networks versus glassy systems.
\newblock {\em Journal of Statistical Mechanics: Theory and Experiment},
  2019(12):124013, dec 2019.

\bibitem{baityjesi:19b}
M.~Baity-Jesi and V.~Mart\'in-Mayor.
\newblock Precursors of the spin glass transition in three dimensions.
\newblock {\em Journal of Statistical Mechanics: Theory and Experiment},
  2019(8):084016, 2019.

\bibitem{franz:19}
Silvio Franz, Sungmin Hwang, and Pierfrancesco Urbani.
\newblock Jamming in multilayer supervised learning models.
\newblock {\em Physical review letters}, 123(16):160602, 2019.

\bibitem{bahri:20}
Yasaman Bahri, Jonathan Kadmon, Jeffrey Pennington, Sam~S Schoenholz, Jascha
  Sohl-Dickstein, and Surya Ganguli.
\newblock Statistical mechanics of deep learning.
\newblock {\em Annual Review of Condensed Matter Physics}, 11(1), 2020.

\bibitem{marcus:22}
Stav Marcus, Ari~M Turner, and Guy Bunin.
\newblock Local and collective transitions in sparsely-interacting ecological
  communities.
\newblock {\em PLoS computational biology}, 18(7):e1010274, 2022.

\bibitem{newman:96c}
C.~M. Newman and D.~L. Stein.
\newblock Non-mean-field behavior of realistic spin glasses.
\newblock {\em Phys. Rev. Lett.}, 76:515--518, Jan 1996.

\end{thebibliography}

\appendix

\paragraph{References for the Encyclopedia version}
References for further reading:~\cite{binder:86,mezard:87,mydosh:93,castellani:05,martin-mayor:22,Vincent:22, nordblad:16}

\end{document}